\documentclass[
 reprint,
 amsmath,amssymb,
 aps,
]{revtex4-1}
\usepackage{listings}
\usepackage{graphicx}
\usepackage{dcolumn}
\usepackage{bm}
\usepackage{comment}
\usepackage{color}
\def\qed{\hfill $\Box$} 

\newcommand{\argmax}{\mathop{\rm arg~max}\limits}

\newcommand{\lam}{\lambda}
\newcommand{\tlb}{{T_{\rm lb}}}
\newcommand{\tub}{{T_{\rm ub}}}
\newcommand{\lp}{\lambda'}
\newcommand{\tp}{\tau'}
\newcommand{\hq}{\hat{q}}

\DeclareMathOperator\erf{erf}

\newcommand{\geqapprox}{\raisebox{-2.5pt}[7.5pt][0pt]{$\overset{>}{\sim}$} }

\begin{document}

\preprint{APS/123-QED}

\title{When to wake up? The optimal waking-up strategies for starvation-induced persistence}

\author{Yusuke Himeoka}
 \email{yusuke.himeoka@nbi.ku.dk}
\author{Namiko Mitarai}
 \email{mitarai@nbi.ku.dk}
\affiliation{
The Niels Bohr Institute, University of Copenhagen, Blegdamsvej 17, Copenhagen, 2100-DK, Denmark }
\date{\today}
\begin{abstract}
Prolonged lag time can be induced by starvation contributing to the antibiotic tolerance of bacteria. We analyze the optimal lag time to survive and grow the iterative and stochastic application of antibiotics. A simple model shows that the optimal lag time exhibits a discontinuous transition when the severeness of the antibiotic is increased. This suggests the possibility of reducing tolerant bacteria by controlled usage of antibiotics application. When the bacterial populations are able to have two phenotypes with different lag times, the fraction of the second phenotype that has different lag time shows a continuous transition. We then present a generic framework to investigate the optimal lag time distribution for total population fitness for a given distribution of the antibiotic application duration. The obtained optimal distributions have multiple peaks for a wide range of the antibiotic application duration distributions, including the case where the latter is monotonically decreasing. The analysis supports the advantage in evolving multiple, possibly discrete phenotypes in lag time for bacterial long-term fitness.  
\end{abstract}
\maketitle

\section{Introduction.} When bacterial cells are transferred from a starving environment to a substrate-rich condition, it takes sometime before the cells start to grow exponentially. This lag phase \cite{monod1949growth} is often considered as the delay during which the cells modify their gene expression pattern and intracellular composition of macromolecules to adapt to the new environment \cite{himeoka2017theory,madar2013promoter,larsen2006differential,rolfe2012lag,bathke2019comparative}. Therefore, the characteristics of the lag phase depend on the growth condition before the starvation, the environment during the starvation, and the new environment for the regrowth. In spite of this complexity, na\"{\i}vely thinking, reducing the lag time as possible appears to be better for the bacterial species because it maximizes the chances of population increase. However, interestingly, it has been reported that the distribution of the lag time at a single-cell level has much heavier tail than a normal distribution \cite{levin2010automated,levin2016quantitative,Simsek17635}. 

Indeed, having a subpopulation with long lag time can be beneficial under certain circumstances, for instance, when the nutrients are supplied together with antibiotics. This is because antibiotics often target active cellular growth processes and hence dormant, non-growing cells are tolerant of the killing by antibiotics \cite{tuomanen1986rate,eng1991bactericidal,lewis2010persister}. In general, dormant cells tend to be less sensitive to environmental stress, providing a better chance of survival. Therefore, the lag phase can work as a shelter for the cells from lethal stress. 

The phenotypic tolerance provided by a dormant subpopulation has been attracting attention as a course of bacterial persistence \cite{bigger1944treatment,balaban2004bacterial,lewis2010persister,fisher2017persistent,balaban2019definitions}. Operationally persistence can be categorized into two types \cite{balaban2004bacterial,balaban2019definitions}; type I or triggered persistence, where the dormancy is triggered by external stress such as starvation, and type II or spontaneous persistence, where the cells switch to a dormant state even though the environment allows exponential growth of the population. 
The spontaneous persistence has been interpreted as a bet-hedging strategy \cite{kussell2005bacterial,kussell2005phenotypic,gardner2007bacterial,patra2013population}, where the optimal switching rates between dormancy and growth is proportional to the switching rates of the environments with and without antibiotic. 
For the triggered persisters, there should also be an optimal lag time distribution under given antibiotics application. Analysis of the optimal lag time should be relevant in understanding bacterial persistence given a recent laboratory experiment with {\it Escherichia coli} showing that the starvation triggers the dominant fraction of the bacterial persistence \cite{harms2017prophages}, as well as their appearance in pathogenic bacteria {\it Staphylococcus
aureus} and its correlation with antibiotic usage~\cite{vulin2018prolonged}. 

Previously, Friedman {\it et al.} \cite{fridman2014optimization} have conducted an experiment to see whether bacteria can evolve to increase the lag time by an iterated application of the antibiotics at a lethal level. In the experiment, they grew bacteria with fresh media supplemented with an antibiotic. The antibiotic was removed after a fixed time $T$ had passed, and the culture was left for one day to let the survived bacteria grow and enter the stationary phase. Then, a part of the one-day culture was transferred to the next culture, supplemented fresh media with the antibiotic. By repeating the procedure, it was found that the mean lag time of the bacteria has evolved to the roughly same length to $T$, which is expected to be optimal for the long-term population growth. 

The present work is motivated by this experiment. In their experiment, the antibiotics were applied at every re-inoculation. What will happen if the application of the antibiotic is probabilistic and the duration of the antibiotics application fluctuates? What are the optimal distribution and the average of the lag time? Is it better for the total growth to split into subpopulations with different lag times? 

In the first part of the present paper, we analyze the optimal waking-up strategy under the probabilistic antibiotic application by using a simple population dynamics model. Analytical and numerical calculations show that the evolution to increase the lag time occurs only if the effect of the antibiotics exceeds a certain threshold, at which a discontinuous transition of the optimal lag time from zero to finite happens.  We then extend the model so that the population can be split into two subgroups with different average lag time, to show that there is a continuous transition from single-strategy to bet-hedging strategy when changing the antibiotics application probability and time. 

The setup is then generalized in the latter part to ask the optimal lag time distribution without specifying the dynamics in the lag phase and the distribution of the antibiotics application time. The optimal lag time distribution is analytically shown to have a finite gap region in which the probability is zero-valued. Also, the optimal lag time distribution for several distributions of the antibiotics application time is concretely computed. Finally, the implication of the calculated optimal lag time distribution to the biologically observed lag time distribution is discussed. 

\section{Model with a constant rate wake-up.} Motivated by ref.~\cite{fridman2014optimization}, we consider the following setup: Bacterial cells are transferred from stationary phase culture to a fresh media where all the cells are in a dormant state. At every re-inoculation, the fresh media is supplemented with the antibiotics with the probability $p$. The antibiotics are removed at time $T$, and after that, the culture will be left to grow long enough time until $t\gg T$ before entering the stationary phase.

We assume that a cell can take two states, namely, the dormant (or lag) state and the growing state. A cell in the dormant state is assumed to be fully tolerant of the antibiotics but cannot grow (this assumption can be relaxed. See Appendix.\ref{sec:relax_assumption}). A cell in the growing state dies at a rate of $\gamma$ if the antibiotics exist in the environment, and proliferates at a certain rate if there is no antibiotic. Here we suppose that the antibiotics are bactericidal, but not bacteriostatic because the application of the bacteriostatic antibiotics just leads to the prohibition of bacterial activities, and then, there is no meaning in discussing the optimal waking-up strategy.

To be concrete, we first analyze a case where the cells in the dormant state transit to the growing state at a constant rate of $1/\lambda$ as in ref.~\cite{fridman2014optimization}. 
Here $\lambda$ corresponds to the average lag time of the population, 
and the lag time distribution follows an exponential function. 
The transition from a growing state to the dormant state is not considered when there are nutrients in the culture. 
Hereafter, we set the proliferation rate to unity by taking its inverse as unit of time. Then, the temporal evolution of the population after an inoculation is ruled by a linear ordinary differential equations as follows; 
\begin{eqnarray}
\frac{d}{dt}d(t)&=&-d(t)/\lambda,\label{eq:dynamics}\\ 
\frac{d}{dt}g(t)&=&
\begin{cases}
d(t)/\lambda-\gamma g(t) & (t<T)\\
d(t)/\lambda+ g(t) & (t>T),\\
\end{cases}\label{eq:dynamics2}
\end{eqnarray}
where $d$ and $g$ represent the population in the dormant state and growing state, respectively. The population dynamics of the antibiotic-free case is obtained by setting $T=0$. We set the initial population to unity without losing generality, i.e., $d(0)=1,\ g(0)=0$.

Since $d(t)=e^{-t/\lambda}$, in $t\gg\lambda$ region, $g(t)$, being asymptotically equal to $f(T)e^t$, represents the total population at time $t$. By noting that the population with zero-lag time grows as $\exp(t)$ under the antibiotic-free condition, $f(T)=g(t;T)/\exp(t)$ measures the impact of the antibiotics and the cost of having non-zero lag time as the population loss relative to the exponential growth without the antibiotics and the lag time. Since $f(T)$ is the measure for a single round of the inoculation, with many repetition of this process, the long-term average normalized growth per round $F_I(\lambda;\gamma,p,T)$ is given by averaging $\ln[f(T)]$ over the probability $p$ of the antibiotics application \cite{kelly1956new,cohen1966optimizing,maslov2015well,bergstrom2004shannon} as 
\begin{eqnarray}
    &&F_I(\lambda;\gamma,p,T)=(1-p)\ln\Biggl[\frac{1}{1+\lambda}\Biggr] \nonumber\\
    &&+p\Biggl(-T+\ln\Biggl[\frac{e^{-T/\lambda}-e^{-\gamma T}}{\gamma\lambda-1}+\frac{e^{-T/\lambda}}{1+\lambda}\Biggr]\Biggr).\label{eq:fit_M1}
\end{eqnarray}
Hereafter, we study the optimal lag time $\lambda^*$ which maximizes the fitness $F_I$ for a given environmental parameter set $p, T$, and $\gamma$.

\section{Optimal lag time.}
\subsection{Linear-wake up model}

\begin{figure}[tbp]
\begin{center}
\includegraphics[width = 85 mm, angle = 0]{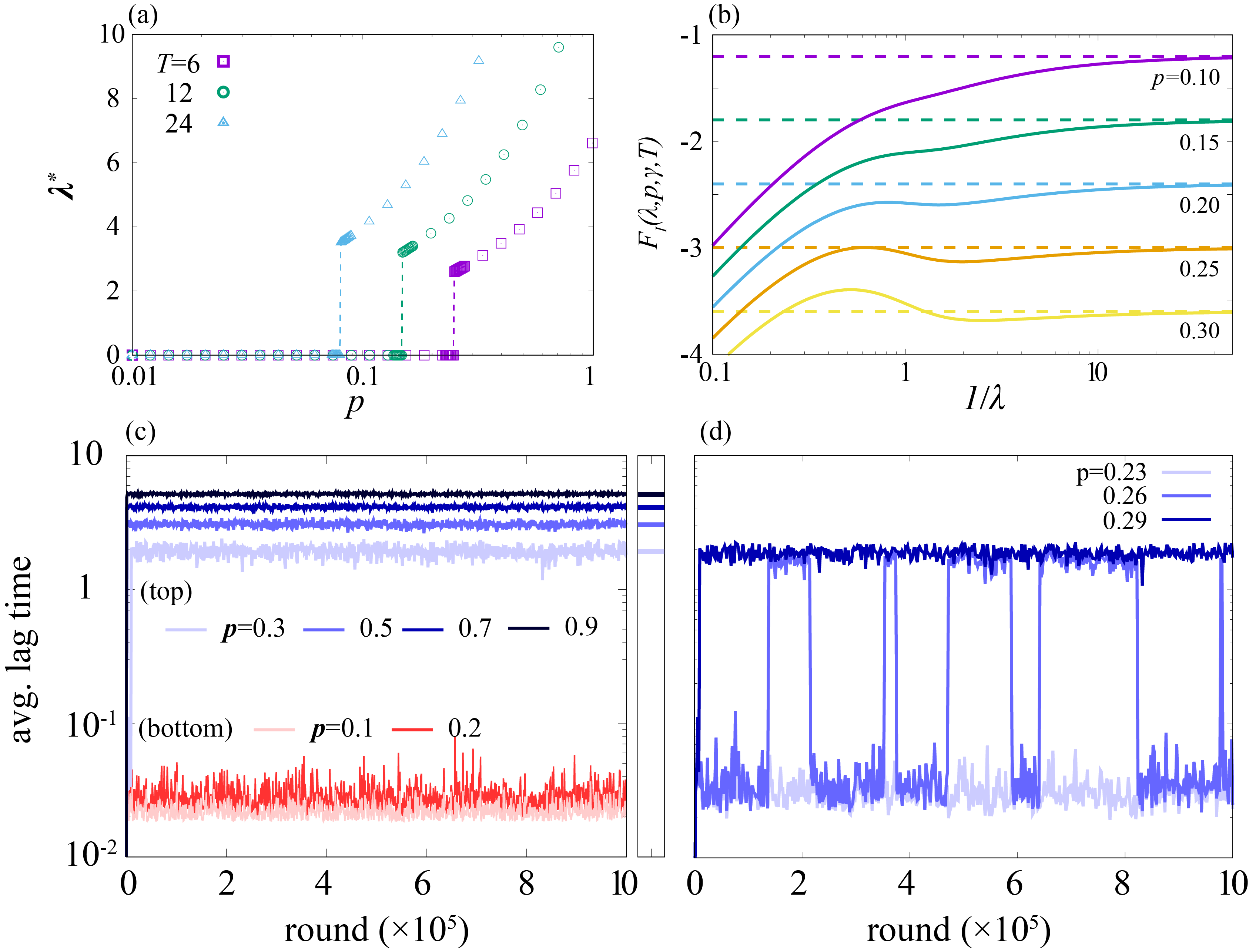}
\caption{(a). The optimal lag time value, $\lambda^*$, is plotted as a function of $p$ for several choices of $T$ value. The optimal lag time shows the discontinuous transition. Below the transition point, $\lambda^*$ is zero.  (b). The fitness function is plotted as a function of $1/\lambda$ for several values of $p$ close to  the critical value. Each dashed line represents $\lim_{1/\lambda\to\infty}F_I(\lambda,\gamma,p,T)$. The local peak of the fitness is formed at $p\approx 0.2$, and the fitness value at the peak exceeds its value at $1/\lambda\rightarrow \infty$. The local optimal of $F_I$ is formed at $p\approx 0.2$, and it becomes the global optimal at $p\approx 0. 25$. (c) the evolutionary time courses of the averaged lag time in the serial selection model are plotted for several values of $p$. The right panel shows the optimal lag time predicted from the optimization of $F_I$ for corresponding colors while it is not plotted for $p=0.1$ and $0.2$ because the optimal value is zero. (d). The evolution time courses are computed around the critical $p$ value. At $p=0.26$, the averaged lag time shows a bistable behavior. The figure shares the y axis with (c). $\gamma$ is set to unity and $T=6$ for (b-d).}
    \label{fig:Onestep}
  \end{center}
\end{figure}
First, we address the case where the duration of the antibiotics application has no fluctuation. In Fig.~\ref{fig:Onestep}(a), the optimal lag time $\lambda^*$ is plotted as a function of $p$ for various values of $T$, with $\gamma=1$. The killing rate $\gamma\approx 1$ is biologically reasonable range since it is often found to be the same order of magnitude of the bacterial growth rate \cite{lee2018robust,tuomanen1986rate}. 
When changing $p$, the optimal lag time $\lambda^*$ shows a discontinuous transition, with a critical $p$ value dependent on $T$ (and $\gamma$). Below each critical $p$ value, the optimal lag time is zero, while above the threshold, it increases with $p$ as $\lambda^*\sim pT$, reaching $\lambda^* \sim T$ when antibiotic is always present ($p=1$).  
The fitness $F_I$ is plotted as a function of $1/\lambda$ in Fig.~\ref{fig:Onestep}(b), demonstrating the appearance of a local maximum at a positive finite $\lambda$ leading to a discontinuous transition above a critical $p$. We found that the transition takes place by changing the severeness of the application of the antibiotic, by changing one of the parameters among $\gamma,p$, and $T$ with keeping the rest constant. In the $\gamma\to\infty$ limit, it is easy to show that the optimal lag time is given by $1/\lambda^*=(-1+\sqrt{1+4/pT})/2$. For the killing rate $\gamma$ and the probability of the antibiotics application $p$, we give proof for the existence of the critical values at which the transition occurs and an upper bound of the critical $\gamma$ being $1/p-1$ in {\it Supplementary Information} section.$I$.

Note that, in this setup, the total population is allowed to be infinitesimally small. However, in reality, the population size less than a single cell means extinction.   
In order to take the discreteness of the number of the cells into account, we also performed an optimization of the fitness with a constraint $(d(t)+g(t))\geq\delta_{\rm ext}$ where $\delta_{\rm ext}$ represents the population allocated for a single cell in our unit. This modification makes another continuous transition of the optimal lag time from zero to non-zero when $p$ and $T$ are varied, but the model still exhibits the discontinuous transition, too (see Appendix.\ref{sec:extinction}). 

\subsection{Comparison with sequential selection procedure}
It is worth mentioning the meaning of the optimization of $F_I$. The fitness function $F_I$ is defined as the average of the logarithmic growth over the multiple rounds of waking-up experiment. Thus, the optimization of $F_I$ operationally corresponds to picking up the bacterial culture that grows most successfully after multiple rounds of inoculation among a number of parallel cultures where cells in each parallel culture have different values of $\lambda$. This process is different from an inoculation-dilution cycle performed in ref.~\cite{fridman2014optimization} where a variety of phenotypes can exist and fitters will be selected by the frequency-dependent selection at the stage of the dilution. 

In order to ask if the cycle results in different consequences from what we have found by optimizing $F_I$, we performed the inoculation-dilution simulation.
The setup of the simulation is the following; we reserve $2N$ variables for the populations of $N$ types of the cells, $d_i$ and $g_i\ (0\leq i<N)$, who has the lag time $\lambda_i$ ($\lambda_i<\lambda_{i+1}$). Our model cells are incubated {\it in silico} until $t=\tau\ (\tau>T)$ from the initial state that all the cells are in the dormant state. The cells of the $i$th type wake up at the rate $1/\lambda_i$. The waken-up cells are killed if the antibiotics are applied and $t<T$, otherwise proliferates. When $t$ reaches to $\tau$, the cells are harvested and diluted. 

To make the evolution of the lag time possible, we introduced a mutation to the model. When a single cell of the $i$th type divides, the daughter cell can mutate to the $(i-1)$th or the $(i+1)$th type who has a slightly different lag time. Overall, the temporal evolution of the population is ruled by the linear equation (Eqs.~(\ref{eq:dynamics})-(\ref{eq:dynamics2})) with a replacement of $\lambda$ by $\lambda_i$ and addition of the mutation term under the growing condition (the detailed equations are provided in Appendix.\ref{sec:evolution}). 

In the dilution process, each type of the cells is diluted proportionally to its fraction in the harvested culture \footnote{i.e, $d_i(0)$ for the next round is given as $(d_i(\tau)+g_i(\tau))/\sum_j(d_j(\tau)+g_j(\tau))$  of the current round, and $g_i(0)$  for the next round is zero. }, and thus, it leads to the frequency-dependent selection. Since all the parameters are the same among $i'$s except the lag time, the cells with an adequate lag time is supposed to fit this sequential culture the most and to be selected. We introduced the smallest number of the cell $\delta_{\rm ext}$ also to this simulation. Every time the antibiotics application ends and the dilution is completed, we check the number of the cells of each phenotype and if its value is below the threshold, the value is truncated to zero.

Also, to be consistent with the present model, the antibiotic is applied in probability $p$ and for the duration of $T$. The incubation time $\tau$ is set significantly longer than the studied values of antibiotics application duration $T$ so that there is enough time for exponential growth. Concretely, we studied range of $T$ being from $0.0$ to $6.0$ and used $\tau=20.0$. 

Fig.~\ref{fig:Onestep}(c) and (d) show the evolutionary time course of the average lag time $\langle \lambda \rangle$ over the population for several values of $p$. The evolution simulation was initiated from all the population that has the shortest lag time (i.e, $d_0(0)=1$). As shown in Fig.~\ref{fig:Onestep}(c), for the small $p$ values, the population stays at $\langle \lambda \rangle < 10^{-1}$ which corresponds to having the peak at $i=0$ in the population distribution in $i$ space. In contrast, for $p\geq 0.3$, the average lag time increases over the rounds and settles down at a certain value which is consistent with the value predicted by the optimization of $F_I$. Around the critical $p$ value, the evolutionary dynamics showed a bistable behavior as shown in Fig.~\ref{fig:Onestep}(d). The critical $p$ value here is inferred as around $0.26$ being reasonably close to the critical $p\approx 0.25$ in Fig.~\ref{fig:Onestep}(b).

\subsection{Distributed antibiotic application time}
\begin{figure}[tbp]
\begin{center}
\includegraphics[width = 85 mm, angle = 0]{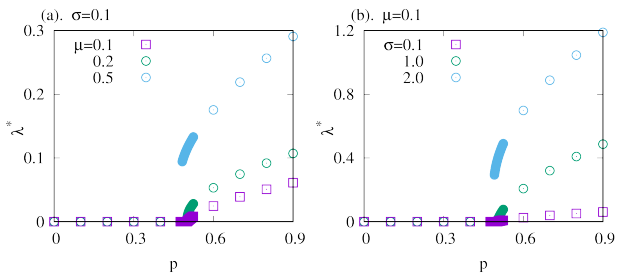}
\caption{The optimal average lag time $\lambda^*$ is plotted as a function the probability of the antibiotics application $p$. A normal distribution with its mean $\mu$ and the standards deviation $\sigma$ is used for the distribution of the antibiotics application time $q(T)$. The transition gets more and more steeper as either (a). the average $\mu$ or (b). the standard deviation $\sigma$ increases. $\gamma$ is set to unity.}
    \label{fig:continuous}
  \end{center}
\end{figure}
The discontinuous transition of the optimal lag time observed so far is easy to interpret as a result of competition between the zero lag time being optimal for the no-antibiotic case, while $T$ being optimal when antibiotic is applied. A nontrivial question is then if the transition stays discontinuous when the antibiotic application time $T$ is distributed. Therefore, we generalized the analysis to the case where the antibiotic application time $T$ fluctuates. The fitness function for such cases is defined by replacing the probability of antibiotics application $p$ in Eq.~(\ref{eq:fit_M1}) by $pq(T)$ where $q(T)$ is the probability distribution function of the duration $T$. 

The optimal lag time with a normal distribution as $q(T)$ is plotted against the total probability antibiotics application $p$ in Fig.~\ref{fig:continuous}. The transition becomes more and more gradual as the peak of $q(T)$ approaches to zero or $q(T)$ becomes broader. Indeed, the lower bound of the support of $q(T)$ to be non-zero is found as a sufficient condition of the discontinuous transition of the optimal lag time in the model Eq.(\ref{eq:dynamics}) regardless of the choice of $q(T)$ (see {\it Supplementary Information} section.II).

\subsection{Multi-step sequential wakeup model}
Finally, we have performed an analysis of an extended version of the  model (Eq.(\ref{eq:dynamics})-(\ref{eq:dynamics2})) where the cells sequentially go through multiple ($M$) dormant states as follows:
\begin{eqnarray}
\frac{d}{dt}d_0(t)&=&-d_0(t)\cdot M/\lambda,\label{eq:multimodel0}\\ 
\frac{d}{dt}d_i(t)&=&(d_{i-1}-d_i(t))\cdot M/\lambda, \ (1\leq i \le M-1) \label{eq:multimodeli}\\ 
\frac{d}{dt}g(t)&=&
\begin{cases}
d_{M-1}(t)\cdot M/\lambda-\gamma g(t) & (t<T)\\
d_{M-1}(t)\cdot M/\lambda+ g(t) & (t>T),\\
\end{cases}\label{eq:multimodelg}
\end{eqnarray}
where $M/\lambda$ is the rate of the transition from the $i$th to the $i+1$th state. Note that $M=1$ corresponds to the original model (Eq.(\ref{eq:dynamics})-(\ref{eq:dynamics2})).
The introduction of the multiple steps leads to the Erlang distribution with an average $\lambda$ as the lag time distribution. In Appendix.\ref{sec:multistep}, we showed that the discontinuous transition is triggered also in the extended model. The more dormant states the system has, the better the fitness function gets at its optimal $\lambda$ under relatively large $p$ because the distribution is narrower.\\
Also, the same argument with the single-step model (Eq.(\ref{eq:dynamics})-(\ref{eq:dynamics2})) on the discontinuous transition holds for arbitrary numbers of the intermediate dormant states, i.e., the non-zero lower bound is the sufficient condition for $M$-step sequential models to exhibit the discontinuous transition in the lag time, regardless of the choice of $q(T)$ (see {\it Supplementary Information} section.II).

\section{Bet-hedging.} So far, we have studied the optimal lag time where all the cells have the same transition rate ($1/\lambda$). However, it is known that even an isogenic bacterial population can split the population into several phenotypes. To see whether the best strategy changes in the multi-phenotype case, we did the simplest extension of the single-step model (Eq.(\ref{eq:dynamics})-(\ref{eq:dynamics2})) to the case where the bacteria is capable of having two subpopulations with different values of average lag time. 

We split the total population into two parts, $a$ and $b$, and assume the transition rates from the dormant to growing state being $1/\lambda_a$ and $1/\lambda_b$, respectively. Without loss of generality, we assume $\lambda_a\le \lambda_b$, and denote the fraction of the subpopulation $b$ as $x$. The fitness function is then written down as
\begin{eqnarray}
&&F_{I\hspace{-0.8mm}I}(x,\lambda_a,\lambda_b;\gamma,p,T)=p\ln\biggl[(1-x) f_a(T)+x f_b(T)\biggr]\nonumber \\
&&+(1-p)\ln\biggl[(1-x)f_a(0)+xf_b(0)\biggr],
\end{eqnarray}
where $f_a(T)$ ($f_b(T)$) is $\ln[g(t;T)/\exp(t)]$ with the transition rate $1/\lambda_a$  ($1/\lambda_b$). In this case, the parameters that the bacteria can evolve to optimize are $\lambda_a$, $\lambda_b$, and $x$. 

\begin{figure}[tbp]
\begin{center}
\includegraphics[width = 85 mm, angle = 0]{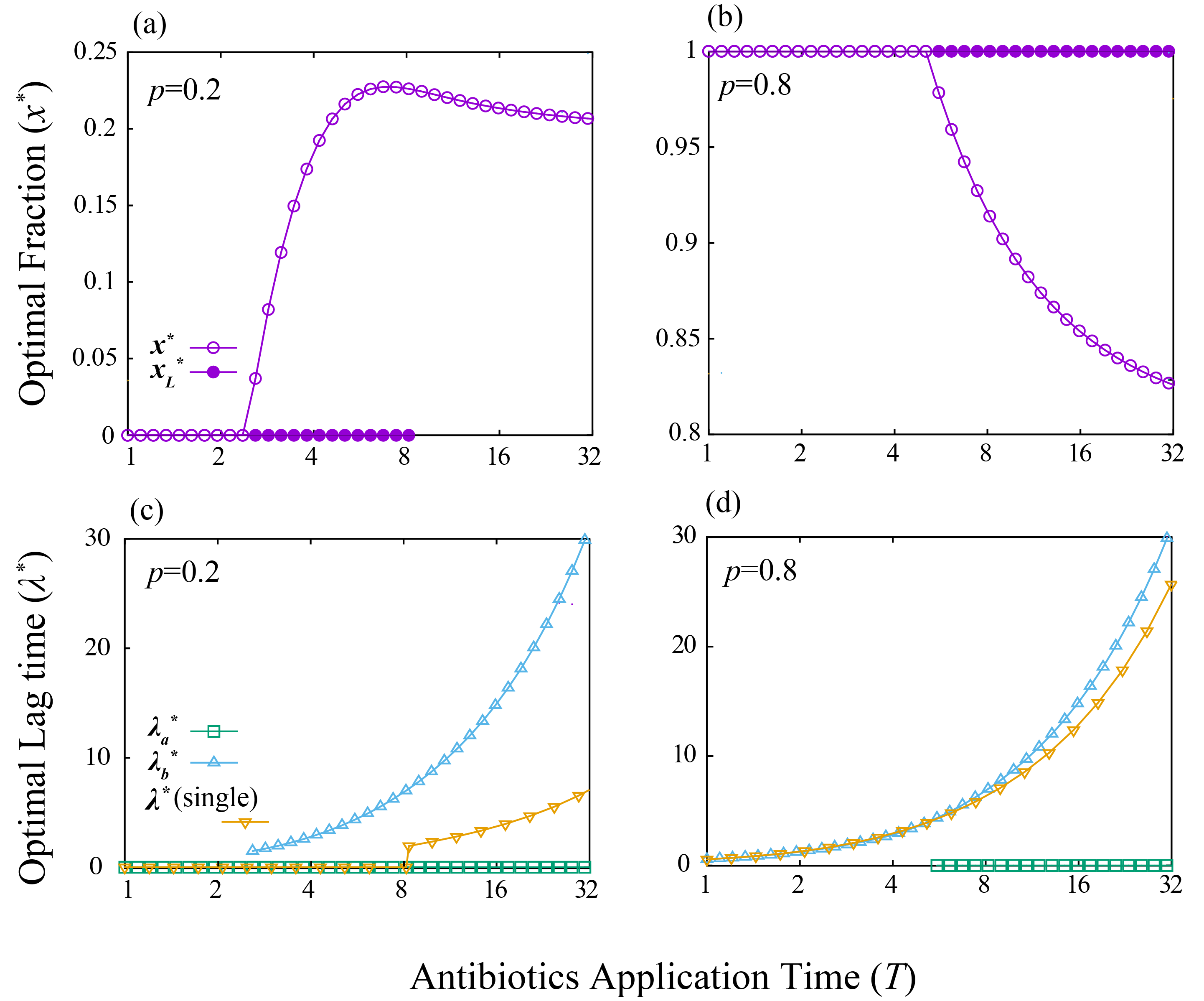}
\caption{ The optimal fraction of the strategy with non-zero lag time $x^*$ ((a) and (b)) and  the optimal lag times of both two strategies ($\lambda^*_a=0$ and $\lambda^*_b$) and the one strategy case ($\lambda^*$) ((c) and (d)) are plotted against the antibiotics application time $T$ for different probability of the antibiotics application $p$. In (c) and (d), the optimal lag time of the phenotype is plotted only if its globally fraction in the population is non-zero because any value is allowed as the optimal if the fraction is zero. $p=0.2$ for (a) and (c), and $0.8$ for (b) and (d). $\gamma$ is set to unity. }
    \label{fig:Twosp}
  \end{center}
\end{figure}
Fig.~\ref{fig:Twosp} shows the optimal parameter values as a function of the antibiotics application time $T$ for different probability $p$, obtained by the numerical optimization of $\lambda_a$, $\lambda_b$, and $x$. The optimal lag time of the single-strategy case is also shown for comparison. 

Interestingly, the probability of the antibiotics application changes the qualitative behavior of the optimal fraction ($x^*$) to an increase of $T$. For a small $p$ value (Fig.~\ref{fig:Twosp}(a) and (c)), all the cells have zero lag time in the short $T$ region ($x^*=0$), while as $T$ gets longer, the population start to invest subpopulation to the non-zero lag time phenotype. This strategy corresponds to the bed-hedging strategy which is investigated by Kelly as an extension of the information theory to gambling \cite{kelly1956new}, and later, studied widely for instance, in the population dynamics field \cite{cohen1966optimizing,maslov2015well} including type-II persistence \cite{kussell2005phenotypic,kussell2005bacterial} as well as the finance \cite{maslov1998optimal} and the relation between the information theory and biological fitness \cite{bergstrom2004shannon}; 
as the risk of the antibiotic application becomes larger with longer $T$, it pays off to save a small fraction of the population to finite lag time to hedge the risk.  
On the other hand, if the value of $p$ is relatively large (Fig.~\ref{fig:Twosp}(b) and (d)), all the population has non-zero lag time ($x^*=1$) even if $T$ is small, reflecting that the chance of having antibiotics is too high that it does not worth betting subpopulation into zero lag time going for more growth in no-antibiotic condition. However,  as $T$ gets larger, the optimal strategy changes to bet some fraction of the population to zero-lag time. This somewhat counterintuitive result is due to the trade-off between the benefit and cost of having a longer lag time. Cells can avoid getting killed by having the lag time being close to the antibiotics application time $T$. However, having a long lag time means waiving the opportunity to grow even when the fresh media is fortunately antibiotics-free. While the loss of the opportunity is negligible for small $T$, as $T$ gets larger, the loss becomes sizable and it becomes better for the population to bet some fraction for the chance of the media to be antibiotics-free.

In the analysis, we also found some locally optimal strategies, which are shown in Fig.~\ref{fig:Twosp} (a) and (b) as $x^*_L$. For both $p=0.2$ and $p=0.8$, the optimal strategy for short $T$ is having a single phenotype ($x^*=0$ for (a) and $x^*=1$ for (b)). For $p=0.2$ case, the single phenotype strategy is locally stable as long as the zero-lag time is the optimal lag time for that strategy ($x^*_L=0$), while after the optimal lag time for single phenotype case $\lambda^*$ becomes non-zero, the single-strategy is no longer optimal, even locally. On the other hand, for $p=0.8$ case, the single phenotype strategy remains locally optimal after the transition in a whole range shown in the figure.

\begin{figure}[tbp]
\begin{center}
\includegraphics[width = 70 mm, angle = 0]{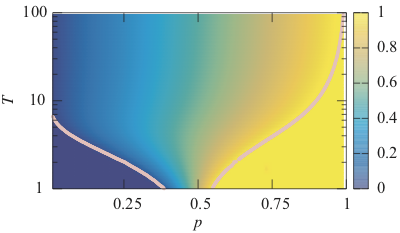}
\caption{The phase diagram of the optimal fraction $x^*$. The solid lines indicate the boundary between $x^*=0$ and $x^*>0$ (left region), $x^*=1$ (right region) and $x^*<1$. $\gamma$ is set to unity. }
    \label{fig:heatmap}
  \end{center}
\end{figure}
In Fig.~\ref{fig:heatmap}, the globally optimal fraction $x^*$ of the subpopulation with finite lag time is shown as a heat-map as functions of $p$ and $T$. There are three phases, namely, (i) all the cells waking up immediately ($x^*=0$), (ii) all the cells have the finite lag time ($x^*=1$), and (iii) the bet-hedging phase ($0<x^*<1$). The $x^*=0$ phase and the $x^*=1$ phase are placed in the region with small $p$ and $T$ and the region with large $p$ and small $T$, respectively. For the shown case of $\gamma=1$, behavior of $x^*$ to the increase of $T$ changes at $p\approx 0.5$. Increasing (decreasing) $\gamma$ shifts the phase boundaries to smaller (larger) $p$ and $T$.

\section{Optimal lag time distribution }
\subsection{Generalized set up}
So far, we have studied the optimal average lag time of the bacterial population with a single and two phenotypes by assuming a specific wake up process. The discontinuous transition of the average lag time and the bifurcation of the strategies are shown to be triggered when the effect of the antibiotics application gets severer. 

However, the model setup itself strongly restricts the possible strategies. For instance, the present sequential model results only in the Erlang distribution as the lag time distribution where the average and the variance are tightly interconnected and the number of phenotypes equals the number of possible peaks. We, however, do not know the detailed wake-up dynamics or the possible number of phenotypes in real biological systems, hence it is hard to know if the imposed restrictions were reasonable. 

Therefore, in this section, we develop a general way to calculate the optimal lag time distribution for a given $T$ distribution, regardless of the internal dynamics of the waking-up, number of phenotypes, and so on. The biological reality will be taken into account afterward to discuss how closer to the optimal distribution the bacterial lag time distribution can approach.

We keep using the same {\it in silico} experimental setup. Instead of introducing a specific population dynamics model, we consider the lag time distribution $r(l)$. Namely, $r(l)dl$ gives the probability for a cell transit from the dormant states to the growing state at time between $t=l$ and $t=l+dl$. There is no growth or death in the dormant states, while in the growing state, the cells grow at the rate 1 in the no-antibiotic environment, or die at the rate $\gamma$ if the antibiotic is applied. The central assumption is that there is only one growing state and it is an absorbing state: a cell cannot go back to the dormant states once it enters the growing state during one cycle. In other words, $r(l)$ is the first passage time distribution to the growing state in one cycle.

First, we formulate the fitness function. If there is no antibiotic applied, an individual with its lag time $l$ grows after time $t=l$, and thus, the dynamics of the total population without the antibiotics at time $t$, $N_0(t)$ is given as
\begin{equation}
N_0(t)[r]=\int_{0}^{t} e^{t-l}r(l)dl+\int_{t}^{\infty}r(l)dl, \label{eq:N0}
\end{equation}
where we assume that the total population is unity at $t=0$. The first term and the second term represents the cells in the growing and dormant states, respectively.

When the antibiotic is applied for a duration $T$, the cells in the growing state die at the rate $\gamma$ during the duration. The total population $N_T(t)$ at time $t>T$ is then given by 
\begin{eqnarray}
N_T(t)[r]
&=&e^t\Bigl(e^{-T}\int_0^Tr(s)e^{-\gamma(T-s)}ds+\int_T^te^{-l}r(l)dl\Bigr) \nonumber\\
&+&\int_t^\infty r(l)dl. \label{eq:Nt}
\end{eqnarray}

Following the earlier argument, $f(T)[r]=\lim_{t\to\infty}\ln[N_T(t)[r]/e^t]$ represents the relative growth of the population having the lag time distribution $r$ under the antibiotics application with the duration $T$. Therefore, the fitness of the population with repeated cycle of starvation and stochastic antibiotic application is given by averaging $f(T)[r]$ over the probability of the antibiotics application itself ($p$)  and the duration ($q(T)$) as
\begin{eqnarray}
    &&F[r,q](\gamma,p)=(1-p)\ln\Biggl[\int_{0}^{\infty} e^{-l}r(l)dl\Biggr]\label{eq:fitness_general}\\
    &+&p\int_0^\infty q(T)\ln\Biggl[e^{-(1+\gamma)T}\int_0^Te^{\gamma l}r(l)dl+\int_T^\infty e^{-l}r(l)dl\Biggr]dT.\nonumber
\end{eqnarray}
This provides the generic definition of the fitness function. Indeed, it results in Eq.~(\ref{eq:fit_M1}) with $T=T_0$ when we chose 
$r(l)=\lambda e^{-l/\lambda}$ and $q(T)=\delta(T-T_0)$.
The optimal lag time distribution is obtained by finding $r(l)$ that maximizes the fitness function ~(\ref{eq:fitness_general}). 

\subsection{General form of the optimal lag time distribution}
Interestingly, it is possible to prove (the detail in {\it Supplementary Information} section III) that the optimal lag time distribution takes the form 
\[r^*(l)=\alpha\delta(l)+(1-\alpha)s(l).
\]
Here, $\alpha$ ($0\leq \alpha\leq 1$) is a continuous function of $\gamma,p$, and $q(T)$. Furthermore, minsupp$(s(l))$ is greater than zero, i.e, there is always a region next to the origin in which the optimal lag time distribution is zero-valued regardless of the choice of $q(T)$. 
Remembering that the optimal lag time distribution with $p=0$ (no antibiotic application) is always given by $\alpha=1$,  the existence of the gap between the origin and minsupp$(s(l))$ means that another peak of the optimal distribution never stems from the delta peak at the origin, but discontinuously appears as $p,\gamma$ or $q(T)$ changes. On the other hand, $\alpha$ (the ratio of the delta peak and other peak(s)) continuously changes with the impact of the antibiotics application. 
When $0<\alpha<1$, $r^*(l)$ has at least two disconnected peaks, which indicates that having at least two distinguishable phenotypes is the optimal strategy. This corresponds to the bed-hedging strategy, where fraction $\alpha$ bets on the no-antibiotic environment to maximize the duration of growth, while the fraction $1-\alpha$ is hedging to survive the antibiotic application period better. 

\subsection{Computing the optimal lag time distribution}
We now summarize a numerical method to obtain the optimal lag time distribution for a given $q(T)$. Here we compute the optimal lag time distribution for $q(T)$ with its upper bound of the support as $T_{\rm max}$ \footnote{Note that the different choices of $T_{\rm max}$ leads to the different distribution of the antibiotics application time, and accordingly, different optimal lag time distribution. For the comparison of the optimal distribution for the different $T_{\rm max}$ values, see Fig.\ref{fig:Tmax}.}. Then, the upper bound of the support of the optimal lag time is shown also to be $T_{\rm max}$ (see {\it Supplementary Information} section III)). Next, we approximate the fitness $F$ by discretizing $r(l)$ and $q(T)$ by bins with the size $\Delta$. The approximated fitness function is given as
\begin{eqnarray*}
&&F_d({\bf r},{\bf q},\gamma,p;N,\Delta)=
(1-p)\ln\Biggl[\sum_{n=0}^{N-1} e^{-n\Delta }r_n\Biggr]\\
&+&p\sum_{m=0}^{N-1}q_m\ln\Biggl[e^{-(1+\gamma)m\Delta }\sum_{n=0}^{m-1}e^{\gamma n\Delta }r_n+\sum_{n=m}^{N-1}e^{-n\Delta }r_n\Biggr],\nonumber
\end{eqnarray*}
where $N=T_{\rm max}/\Delta$. Here, ${\bf r}$ and ${\bf q}$ is the vector of the discretized probability distribution function with the $n$-th element being $r_n=\int_{n\Delta }^{(n+1)\Delta }r(l)dl/\int_{0}^{T_{\rm max}}r(l)dl$, and $q_n=\int_{n\Delta }^{(n+1)\Delta }q(T)dT/\int_{0}^{T_{\rm max}}q(T)dT$, respectively. Note that $\Delta\to 0$ limit leads to $F_d\to F$. The detailed description about the discritization is provided in {\it Supplementary Information} section III. 

The partial derivatives $\{\partial F_d/\partial r_n\}_{n=0}^{N-1}$ converges to the functional variation $\delta F/\delta r$ by taking $\Delta \to 0$ limit because $F_d$ converges to $F$ under this limit and $F$ is the $C^\infty$ functional of $r$. Thus, in principle, one can calculate the optimal lag time distribution $r$ by solving $\partial F_d/\partial r_n=0$ with given value of $\Delta$. 

Together with the constraints $\sum_{n=0}^{N-1}r_n=1$ and $r_n\geq0$ represented by the Karush–Kuhn–Tucker (KKT) multiplier terms \cite{kuhn2014nonlinear,karush2014minima}, the KKT conditions to determine the optimal distribution $\{r_n\}_{n=0}^{N-1}$ is

\begin{eqnarray}
1-\mu_n=\sum_{m=0}^{N-1}q'_m\frac{h_n^m}{\langle h^m\rangle},\ (\mu_n=0\ {\rm or}\ r_n=0) \label{eq:dfdr}
\end{eqnarray}
with $r_n\geq 0$ and $\mu_n\geq 0$, where $\mu_n$ is the KKT multiplier for $r_n\geq 0$. The value of the multiplier for the condition $\sum_{n=0}^{N-1}r_n=1$ is already fixed in the above expression (see {\it Supplementary Information} section III)). 
We here introduced a notation $q'_0=(1-p)q_0$ and $q'_n=pq_n\ (n\geq1)$.  $h_n^m$ is given by 
\begin{equation}
h_n^m=
\begin{cases}
\exp[-m\Delta-\gamma((m-n)\Delta)]\ &(n<m)\\
\exp[-n\Delta]\ &({\rm otherwise}),
\end{cases}
\end{equation}
and $\langle h^m \rangle$ represents the average of $\{h^m_n\}_{n=0}^{N-1}$ over $\{r_n\}_{n=0}^{N-1}$. The $n$th bin has a non-zero value only if the $n$th equation of Eq.~(\ref{eq:dfdr}) is satisfiable with $\mu_n=0$, and otherwise, $r_n$ needs to be zero. Thus, the number of non-zero bins is the same with the number of satisfiable equations with $\mu_n=0$.

Interestingly, the distribution $\{r_n\}_{n=0}^{N-1}$ appears in Eq.~(\ref{eq:dfdr}) only in the form of the average of $h'$s.  Therefore, the number of free variables equals the number of averages in the equation. For instance, if $q(T)$ is the Dirac's delta function $\delta(T-T_0)$ with $T_0>0$ (i.e., fixed $T$ case), only $q'_0$ and $q'_a$ ($a\leq T_0/\Delta<a+1)$) are non-zero. Thus, there are only two free variables; only two bins, say $r_i$ and $r_j$, can be non-zero valued, while others are zero because the number of satisfiable independent linear equations equals to the number of independent variables (note that the equation is linear by regarding $1/\langle h^m\rangle'$s as the variables.). Thus, the fixed $T$ leads to the sum of two Dirac's delta functions as the optimal lag time distribution. 

\begin{figure}[tbp]
\begin{center}
\includegraphics[width = 100 mm, angle = 0]{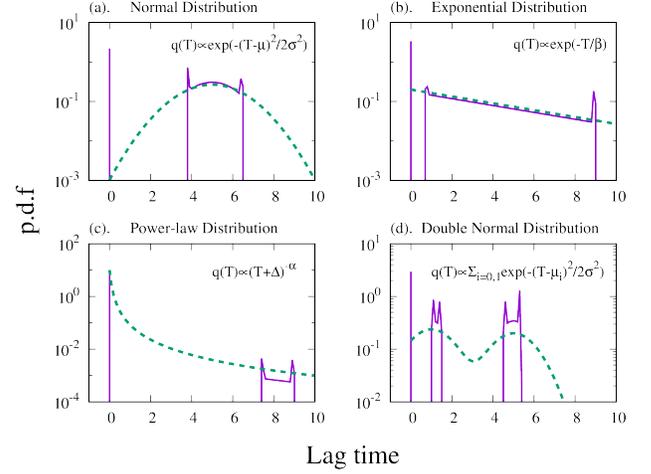}
\caption{The optimal lag time distributions (purple solid lines) obtained by solving Eq.~(\ref{eq:dfdr}) are plotted. The distribution function $q(T)$ (green dashed lines) is (a) a normal distribution with the average $5$ and standard deviation $1.5$, (b) an exponential distribution with the average $5$, (c) a power-law distribution with saturation, $q(T)\propto (T+\Delta)^{-\alpha}$, where $\alpha$ is $2$, and (d) a sum of two equal-weighted normal distributions with the average $1$ and $5$ while the standard deviation is commonly $1.0$. Parameters used in the computation are $\Delta=0.1, T_{\rm max}=10, p=0.8$ and $\gamma=1.0$.}
    \label{fig:opt_dist}
  \end{center}
\end{figure}

By numerically solving the KKT conditions (Eq.~(\ref{eq:dfdr})) for several choices of $q(T)$, the optimal lag time distributions can be obtained. Fig.~\ref{fig:opt_dist} shows the obtained optimal lag time distributions for (a) a normal distribution, (b) an exponential distribution, (c) a power-law distribution, and (d) a sum of two normal distributions as $q(T)$ (the protocol for the computation is described  {\it Supplementary Information} section III). 

The obtained optimal distribution contains a common feature: they consist of the Dirac's delta function-type peak at the origin and the other part mimicking $q(T)$ with steep peaks at its both sides. Interestingly, the optimal lag time distribution has not only the gap next to the origin as proven but also cut-offs in the upper limit. The "mimicking part" can be further divided into the sum of multiple disconnected functions, and the number of the disconnected regions seems to be the same with the number of peaks of $q(T)$ as long as the peaks are distant to each other.

\section{Comparison of the optimal lag time distribution and the results using the multi-step sequential wake-up model}
Lastly, we compare the obtained optimal lag time distribution with the optimal lag time distribution achieved by the $M$-step sequential model (\ref{eq:multimodel0})-(\ref{eq:multimodelg}) extended to multiple number ($N_p$) of phenotypes (details are described in Appendix.\ref{sec:opt_Erlang}). The lag time distribution obtained in this model is always the summation of $N_p$ Erlang distributions.  For a given $M$ and $N_p$, we calculate the fitness value for an Erlang distribution with given average lag time and its fraction in the population for each phenotype by numerical integration to find the optimal lag time values that give the largest fitness value.

\begin{figure}[tbp]
\begin{center}
\includegraphics[width = 95 mm, angle = 0]{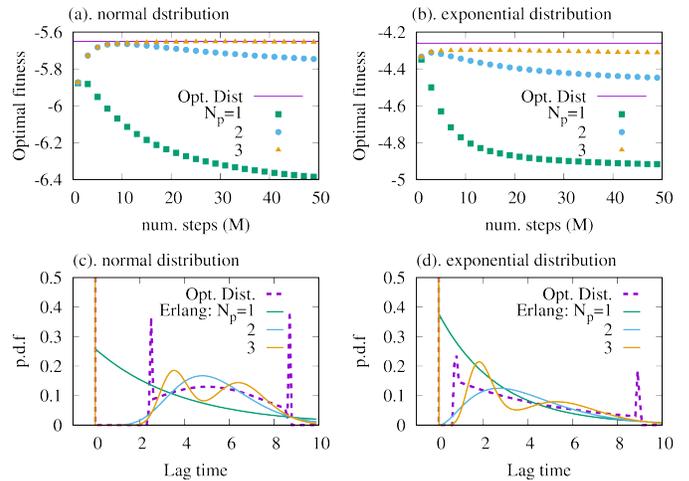}
\caption{
The optimal transition rates are computed for given numbers of subpopulations (types) and states for a normal distribution and an exponential distribution as $q(T)$ ((b) and (d)). The distribution led by the optimal number of states is plotted for each number of types with the optimal lag time distribution computed from Eq.~(\ref{eq:dfdr}) as a reference ((a) and (c)). Note that all the distributions with more than one types a delta function-type peak at the origin. The average and the standard deviation of the normal distribution is $5$ and $3$, respectively, while the parameter for the exponential distribution is $5.0$. The other parameters are given as $T_{\rm max}=10.0, \Delta=0.1, p=0.8$ and $\gamma=10$
}
    \label{fig:opt_erlang}
  \end{center}
\end{figure}

Fig.~\ref{fig:opt_erlang} (a) and (b) shows the largest fitness value obtained by optimized $M$-state $N_p$ phenotype sequential model as function of $M$ for $N_p=1, 2$, and $3$, with $q(T)$ being a normal distribution and an exponential distribution, respectively. The largest fitness value achieved by the solution of Eq.~(\ref{eq:dfdr}) for given $q(T)$ is also shown in the figure. The optimal fitness value is a non-monotonic function of the number of states $M$, while it is increasing function of the number of phenotypes $N_p$ because a population consisting of $N_p$ types can realize any lag time distribution achieved by a population with $n_p<N_p$ types. Interestingly, in these examples, the optimal fitness reaches very close to the fitness value achieved by the solution of Eq.~(\ref{eq:dfdr}) when there are just two phenotypes ($N_p=2$) as long as $M$ is chosen appropriately. 

The obtained optimal lag time distributions for the number of phenotypes $N_p=1,2,$ and $3$ with $M$ chosen to be the best one to maximize the fitness are depicted in Fig.~\ref{fig:opt_erlang} (c) and (d) for the normal distribution and the exponential distribution, respectively. The figure shows that the sequential model with $N_p\ge 2$ is able to successfully capture the essence of the optimal lag time distribution, namely, the delta function at the origin and another peak being distant from the origin. 

\section{Discussion} We have shown that the optimal waking-up strategy changes depending on the severeness of the antibiotic application. When only a single phenotype is allowed, the optimal average lag time exhibits a discontinuous transition from zero lag time to finite lag time as the severeness of the antibiotics is increased. If the cells can split the population to have two subpopulations with different average lag time, bet-hedging behavior can occur depending on the severeness of antibiotics application, where one sub-population has zero lag time and the other sub-population has a positive finite lag time. 

We only presented the analysis of the sequential model for the internal dynamics of the waking-up process, giving the Erlang distribution of lag time for a given phenotype. Another simple case that analysis is straightforward is the case with $\delta$-function distributed lag time, which is presented in {\it Supplementary Information} section IV. There are of course quantitative differences, albeit qualitatively parallel behaviors are obtained in mathematically much clearer form \footnote{In the $\delta$-function case, the optimal lag time is either zero or $T$ because waking up just after $T$ can make the whole population avoid death by the antibiotic. This removes the transition in a single phenotype case with changing $T$.}.

Further, we have developed a generic expression of the fitness function to study the optimal lag time distribution without assuming a specific model for the waking-up dynamics. We have found that the optimal lag time distribution consists of a weighted sum of a delta function at the origin and some functions with a non-zero lower bound of its support. Importantly, there exists a gap between the delta function at the origin and the rest of the functions. This gap could be the origin of the discontinuous transition of the average lag time in the sequential model with a single phenotype. The continuous change of distributions with only a single peak from the delta function at the origin inevitably creates a non-zero valued area inside the gap region because it is not allowed to make another peak to skip the gap region. Since the gap region is one of the worst parts to invest the population, the only way not to waste their "resource" is to reach the area outside the gap region by the discontinuous (fixed $T$) or steep (distributed $T$) transition. 

Of course, the zero-lag time is biologically impossible, hence in reality, what is expected to happen when the zero-lag time is optimal is to evolve to have the shortest possible lag time. The finiteness of the shortest possible lag time affects the location of the critical points as one can infer from Fig.~\ref{fig:Onestep}(b), but as long as it is shorter than the finite optimal lag time after the transition, the qualitative nature of the transition stays the same. Also, the zero-lag time appears in the optimal lag time distribution without any assumption on the wake-up dynamics. From the numerical computations, it is expected that the non-delta function part peaks around the average of the antibiotics application time. Therefore, the gap region could be observed if the shortest possible lag time and the averaged antibiotics application time are distinguishably separated.

Assuming that the bacterial cells evolve to reach the optimal lag time, the present analysis implies the nontrivial evolution of tolerant phenotypes after repeated antibiotics application. If it is easier to evolve the population-averaged lag time as single phenotype than having multiple phenotypes with different lag time, then the discontinuous transition predicted in the single-phenotype case implies the following: If the antibiotics are used very often ($p$ higher than the critical probability for the transition), the treatment leads to the prolonged lag time. In contrast, if the antibiotics are used less often (low $p$), the lag time may shorten as a result of selection, even though $p$ is still non-zero. In other words, there is a critical frequency of antibiotic application below which one can avoid the evolution of more tolerant bacteria. 

Since the transition is sharp for the single phenotype case, it is expected to be relatively easy to detect the transition. If it is easy to evolve to have multiple phenotypes, the total population-averaged lag time would experience a continuous transition because the transition of the fraction $x^*$ is continuous. If the $p$ and $T$ are kept small enough to stay in $x^*=0$ phase, the evolution of tolerant phenotype can be avoided. Therefore, detection of the phase boundary can be clinically important.

It is interesting to note that in standard, single-round inoculation experiments without the antibiotics application, a Gaussian distribution in short lag time with an exponential tail in long lag time \cite{levin2010automated} and a bimodal distribution of lag time \cite{levin2016quantitative} have been reported. The observations suggest that relatively clear division into a few phenotypes can happen. That indicates that the bet-hedging strategy of lag time, which shown to be the best solution for the stochastic antibiotics application at the inoculation, can be realized in the bacterial population.

The present analysis has revealed some interesting differences between type-I or triggered persistence studied here and type-II or spontaneous persistence in terms of the optimization problem. The spontaneous persistence 
has been analyzed as a strategy to cope with the stochastic lethal stress suddenly applied to the environment where cells are already growing exponentially \cite{balaban2004bacterial,kussell2005bacterial,kussell2005phenotypic}. 
When multiple phenotypes that have different growth rates for different environments are allowed, the optimal strategy becomes to mimic the fluctuation of the environment change to switch to the best phenotype. This relatively simple outcome is related to the fact that the growth rate difference is amplified exponentially over time. Therefore, the relation between phenotype switching and environmental switching is simple when we assume that the system stays in an environment for more than a few generation time.  In the present analysis for the type I or triggered persistence, all the cells start at the dormant state, and the difference in lag time provides a difference in the duration to grow (or die) for a finite time (characterized by typical antibiotics application time $T$), but once the antibiotics are removed there is no difference in the growth condition. This subtle difference is still important as we see that in the experimental evolution of lag time \cite{fridman2014optimization}, but the trade-off between waking up too early or too late is relatively small over some range of the lag time. This is the reason for the optimal lag time distributions take a fairly non-trivial shape (Fig.~\ref{fig:opt_dist}). For interested readers, we made a summary of the correspondence and difference between the analysis of optimal strategy in a fluctuating environment by Kussell and Leibler \cite{kussell2005phenotypic} and the present analysis in Appendix.\ref{sec:KL}. 

As a strategy to cope with the fluctuating environment, the responsive adaptation is also a possibility, where the cells sense the environmental fluctuation and respond to it. The benefits and the costs of sensing the environment are very actively investigated for the cells that are already in a growing state \cite{kussell2005phenotypic,xue2018benefits,lan2012energy}. 
During the wake-up from the stationary phase, the cells are sensing the nutrients in the environment and responding to it, as it has been shown in the transcriptomic and proteomic analysis of the temporal patterns of the gene expressions during the lag phase  \cite{larsen2006differential,rolfe2012lag,madar2013promoter,bathke2019comparative}. If the stress factor in the environment that cells are waking up is not so lethal for the cells in the responding-but-not-fully-growing state, it may be possible for a cell in the lag phase to responsively adapt to the stress. It is an interesting future extension to consider the trade-off of sensing and responsive adaptation in the present setup. 

The proposed framework in the present paper provides a way to find the optimal lag time distribution. The actual lag time distribution observed in experiments may be a mixture of the optimality and the restriction imposed on the possible distributions by the physicochemical constraints, history of the evolution, and other factors. Extraction of the pure optimality by the present framework would help future investigations to unveil those restrictions.

Finally, it should be mentioned that the fundamental understanding of bacterial persistence has an impact on various clinical applications. Recently it has been suggested that the persister formation may enhance the emergence of drug-resistant mutant \cite{cohen2013microbial,levin2006non}. Possible mechanisms can be (i) evolution of tolerance supports the rarer resistant mutant to appear \cite{levin2017antibiotic},
(ii) the epistasis between the tolerance and the partial resistance  \cite{levin2017antibiotic},
and/or (iii) enhanced mutation rates \cite{nair2013sub,kohanski2010sublethal,gutierrez2013beta} or horizontal gene transfer \cite{beaber2004sos} in the tolerant cells due to stress response programs. 
Furthermore, accumulating evidence suggests the surprising similarity between the bacterial persistence and antibiotic tolerance of cancer cells \cite{knoechel2014epigenetic,kobayashi2016oct4,fan2011met,raha2014cancer,vinogradova2016inhibitor,hangauer2017drug,sharma2010chromatin,kochanowski2018drug,pearl2016persistence}, and in addition, the triggered persistence-like strategy, escaping from the antibiotics efficacy by staying in the inactive phase, was revealed to be the case for cancer cells \cite{pearl2016persistence}. It has also been argued in the context of the cancer treatment that drug persistent cells can enhance the appearance of drug-resistant cells  \cite{hata2016tumor,ramirez2016diverse,kochanowski2018drug}. The present framework can be used to predict what kind of drug application strategy enhances the appearance of the triggered persistence, which should be taken into account when administrating the drug application.

\paragraph*{Acknowledgement.} This work is supported by the European Research Council (ERC) under the European Union's Horizon 2020 research and innovation programme under grant agreement No. [740704].

\newpage
\setcounter{section}{0}
\begin{widetext}
Note for the {\it arXiv} version: This is the {\it Supplementary Information} part. Appendices follows this part.
\newline

Note : For the consistency with the results shown in the main text, throughout the present document we define the Dirac's delta function peaked at the origin $\delta(x)$ so that $$\int_0^\infty\delta(x)dx=1$$ is satisfied.

\section{The critical $\gamma$ and $p$}
In this section, we describe detailed calculations. For the sake of readability, we first list the notations of some of the functions and parameters.  
\begin{eqnarray*}
F_I(\lam,\gamma,p,T)&=&p\Bigl(-T+\ln\Biggl[\frac{1}{\gamma\lam-1}\Bigl(e^{-T/\lam}-e^{-\gamma T}\Bigr)+\frac{1}{1+\lam}e^{-T/\lam}\Biggr]\Biggr)+(1-p)\ln\Biggl[\frac{1}{1+\lam}\Biggr].\\
F_I^{\infty}(\lam,p,T)&\equiv&\lim_{\gamma\to \infty}F_I(\lam,\gamma,p,T)=-pT(1+\lam^{-1})-\ln[1+\lam].\\
\lam^*(\gamma,p,T)&\equiv&\argmax_\lam F_I(\lam,\gamma,p,T).\\
\lam^*_\infty(p,T)&\equiv&\argmax_\lam F_I^\infty(\lam,p,T)=2\Biggl(-1+\sqrt{1+\frac{4}{pT}}\Biggr)^{-1}
\end{eqnarray*}

First we show there is a critical $0<p<1$. The first-order derivative of the fitness function respect to $\lambda$ is given as
$$\frac{(1+\lam)}{p}\frac{\partial F_I}{\partial \lam}=-\frac{1-p}{p}+(1+\lam)\frac{\partial}{\partial \lam}\ln\Biggl[\frac{1}{\gamma\lam-1}\Bigl(e^{-T/\lam}-e^{-\gamma T}\Bigr)+\frac{1}{1+\lam}e^{-T/\lam}\Biggr].$$
As $p\to0$, the first term of the right hand side of the equation diverges to $-\infty$ while the second term is constant, the optimal $\lam$ is zero. Therefore, the transition of the optimal $\lambda$ triggered by $p$ happens when $(1-p)/p$ approaches to the second term from above to make an intersection. This intersection always exists as long as $\gamma>0$ because 
$$\lim_{\lam\to+0}(1+\lam)\frac{\partial}{\partial \lam}\ln\Biggl[\frac{1}{\gamma\lam-1}\Bigl(e^{-T/\lam}-e^{-\gamma T}\Bigr)+\frac{1}{1+\lam}e^{-T/\lam}\Biggr]=\gamma$$
holds and $(1-p)/p$ ranges in  $(0,\infty)$.  Thus the sufficient condition for the transition to be discontinuous is that the second term is locally an increasing function at the origin of $\lam$. The first order derivative of the second term at the origin is
$$\lim_{\lam\to+0}\frac{\partial}{\partial \lam}(1+\lam)\frac{\partial}{\partial \lam}\ln\Biggl[\frac{1}{\gamma\lam-1}\Bigl(e^{-T/\lam}-e^{-\gamma T}\Bigr)+\frac{1}{1+\lam}e^{-T/\lam}\Biggr]=\gamma(1+\gamma).$$
Therefore, we can conclude that the critical probability $p_c\in(0,1)$ exits and at which the discontinuous transition is triggered. 
\\

The rest of the section is devoted for the critical $\gamma$.  Here we show that there is a value of $\gamma$ at which the optimal $\lam$ value becomes non-zero. First, we see that there is a parameter region in which the optimal $\lam$ is zero. $F_I(\lam,\gamma=0,p,T)$ is a strict monotonically decreasing function of $\lam$, and is a continuous function of $\gamma$ for $0<p<1$ and $T>0$. Thus, for a sufficiently small $\gamma>0$, $F_I(\lam,\gamma,p,T)$ is also the monotonically decreasing function of $\lam$ meaning that the optimal $\lam$ is zero.

To see there is a transition of the optimal $\lam$ value from zero to non-zero, we derive a sufficient condition of $\lam=0$ being no longer optimal. Since $\gamma$ represents the killing rate of the bacterial cells,  $F_I(\lam,\gamma,p,T)\geq F_I^\infty(\lam,p,T)$ and $F_I(\lam^*,\gamma,p,T)\geq F_I^\infty(\lam_\infty^*,p,T)$ hold regardless of the parameter values. Therefore, if $\lam=0$ is the optimal $\lam$ value, 
$$F_I(0,\gamma,p,T)=-pT(1+\gamma)\geq -\frac{pT}{2}\Biggl(1+\sqrt{1+\frac{4}{pT}}\Biggr)-2\coth^{-1}\Biggl(\sqrt{1+\frac{4}{pT}}\Biggr)=F_I^\infty(\lam_\infty^*,p,T)$$
holds. Since the inequality is the necessary condition of $\lam=0$ being the optimal value, the contraposition of this argument is used as the sufficient condition of the transition. 

Note that $F_I(0,\gamma,p,T)$ is a monotonically decreasing function of $\gamma$ with $-\infty$ as its value at $\gamma\to\infty$ limit, whereas $F_I^\infty$ is the constant function. Therefore, there is a value of $\gamma$ ($=\delta$) which satisfies $F_I(0,\delta,p,T)=F_I^{\infty}(\lam_\infty^*,p,T)$. In $\gamma>\delta$ region, $F_I(0,\gamma,p,T)$ is smaller than $F_I^{\infty}(\lam_\infty^*,p,T)$, i.e., $\lam=0$ is no longer the optimal. From the previous argument, one can see that $\delta\neq 0$ holds.

Note that $F_I(\lam^*,\gamma,p,T) > F_I^{\infty}(\lam^*_\infty,p,T)$ always holds for finite $\gamma$ values, and thus, the transition of the optimal $\lambda$ takes place at value of $\gamma\ (=\gamma_c)$ which is strictly lower than $\delta$. 

Next, we derive an upper bound of $\gamma_c$. Note that $\lam^*=0$ is equivalent to that the equation $\partial F_I/\partial (1/\lam)=0$ has no solution of $v\equiv 1/\lam$ in ${\mathrm R}^+$. By taking the partial derivative, we get
\begin{eqnarray}
\frac{\partial F_I}{\partial v}&=&\frac{1}{v}-\frac{1}{1+v}+\frac{p}{\gamma-v}-p\frac{T(1+\gamma)e^{-vT}+e^{-\gamma T}}{(1+\gamma)e^{-vT}-(1+v)e^{-\gamma T}}\\
&=&\frac{1-H(v,\gamma,p,T)}{v(1+v)},\nonumber
\end{eqnarray}
where $H(v,\gamma,p,T)\geq 0$ holds (described later) regardless of parameter values being given as
$$H(v,\gamma,p,T)=\frac{pv(1+v)(1+\gamma)\Bigl[e^{-\gamma T}-\bigl(1-T(\gamma-v)\bigr)e^{-vT}\Bigr]}{(\gamma-v)\Bigl((1+\gamma)e^{-vT}-(1+v)e^{-\gamma T}\Bigr)}$$
Here, $H(0,\gamma,p,T)=0$ and $\lim_{v\to\infty}H(v,\gamma,p,T)=(1+\gamma)p$ holds. Therefore, if $\gamma$ is greater than $1/p-1$, $\partial F_I/\partial v=0$ has a solution. This solution is only the solution if $H$ is a monotonic function of $v$, but from the existence of $\delta$, $H$ always has a solution even if $\gamma$ is smaller than $1/p-1$. Thus,  $\bar{\gamma}_c = 1/p-1$ gives the upper bound of $\delta$ and $\gamma_c$ satisfying $\bar{\gamma}_c\geq\delta>\gamma_c$.

As shown in Fig.\ref{fig:AppB}, $H$ approaches to a monotonic function as $T$ becomes smaller, whereas for large $T$ values, $H$ shows the non-monotonic feature and $\partial F/\partial v=0$ has a solution even if $\gamma$ is smaller than $1/p-1$. 

\begin{figure}[htbp]
\begin{center}
\includegraphics[width = 80 mm, angle = 0]{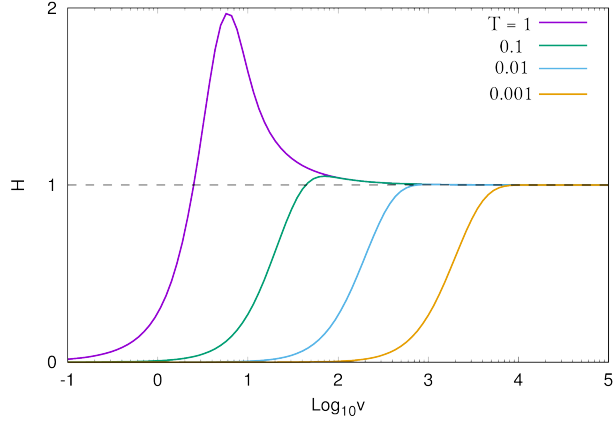}
\caption{$H$ is ploted as a function of $v$. $p=0.2$ and $\gamma=1/p-1$.}
    \label{fig:AppB}
  \end{center}
\end{figure} 

Lastly, we show that $\lambda=0$ has no singularity while the transition is taking place to see the transition is discontinuous. While $\partial F_I/\partial \lambda$ is not continuous at $\lambda=0$, its right limit is given as $p(1+\gamma)-1$. Since $0<\gamma_c < \bar{\gamma}_c$ holds and $p(1+\gamma)-1$ is negative in the region of $\gamma<\bar{\gamma}_c$,  $\lambda=0$ is locally optimal at $\gamma=\gamma_c$ which means that $\lambda^*>0$ is not continuously branching out from $\lambda^*=0$.

$H(v,\gamma,p,T)\geq 0$ is shown as follows; first, the denominator can be rewritten as
$$(\gamma-v)(1+v)e^{-\gamma T}\Biggl(\frac{1+\gamma}{1+v}e^{(\gamma-v)T}-1\Biggr).$$
For the case of $\gamma>v$, $(1+\gamma)/(1+v)>1$ and $e^{(\gamma-v)T}>1$ hold, and thus, the sign of the large parenthesis is positive and $(\gamma-v)>0$ which means the denominator has the positive sign. For the opposite case, $(1+\gamma)/(1+v)e^{(\gamma-v)T}$ is less than one and $(\gamma-v)<0$. Therefore, the denominator is again positive.

The square bracket of the numerator is rewritten as
$$e^{-\gamma T}(1-(1-a)e^a),$$
where $a=(\gamma-v)T$. Here we introduce the function $h(a)$ which is given as $h(a)=1-(1-a)e^a$. By noting that $h(0)=1$ and $dh/da= ae^a$ hold, one can see that $h(a)$ monotonically increases (decreases) with $a$ in the region of $a>0$ ($a<0$),
and $h(0)$ is the minimum. 
Therefore, $h(a)$ is always larger than one, and accordingly, positive.

\section{A condition for a discontinuous transition for arbitrary probability distribution functions}
In this section, we provide a proof for a sufficient condition for the sequential models with distributed antibiotics application time to exhibit the discontinuous transition of the optimal lag time. The sufficient condition is that the lower bound of the support of $q(T)$, the distribution function of the antibiotics application time, is non-zero. For arbitrary functions $q(T)$ satisfying the condition, there exists the critical probability of the antibiotics application, $p_c$ in $(0,1)$ such that the optimal lag time is $0$ for $p<p_c$ while it discontinuously transits at to a non-zero value at $p= p_c$. We denote supp$(q)$ by $(\tlb,\tub)$. While we assume that the support consists of the single connected interval, the following argument is easily extended for cases that the support has more than one connected components.

This claim is shown as following: first, the fitness function is
$${\cal F}_I=(1-p)\ln f(0)+p\langle \ln f(T)\rangle,$$
where $f(T)$ is the single-round fitness of $M$ states model given as
\begin{eqnarray*}
f(T)&=&e^{-(1+\gamma)T}(1-\gamma\lambda)^{-(1+M)}\\
&+&e^{-(1+1/\lambda)T}\sum_{n=1}^{M+1}\frac{(T/\lambda)^{M+1-n}}{(M+1-n)!}\Bigl((1+\lambda)^{-n}-(1-\gamma\lambda)^{-n}\Bigr),
\end{eqnarray*}
and $\langle \cdot \rangle$ means the average over the distribution $q(T)$ (here we dropped the subscript $M$ on $\lambda$ for the readability). Accordingly, the first derivative of the fitness function is given by
$$c\frac{\partial {\cal F}_I}{\partial \lambda}=\frac{\partial \langle \ln f(T)\rangle/\partial \lambda }{\partial \ln f(0)/\partial \lambda}-\Bigl(-\frac{1-p}{p}\Bigr)\equiv {\cal G}_I(\lambda,\gamma)-{\cal H}_I(p),$$
where $1/c=p\cdot \partial\ln f(0)/\partial \lambda$. The non-zero optimal $\lambda$ is determined by ${\cal G}_I(\lambda,\gamma)={\cal H}_I(p)$. Since ${\cal H}_I(p)$ diverges to $-\infty$ as $p\to 0$ and ${\cal G}_I$ is bounded, the transition triggered by $p$ happens when ${\cal H}(p)$ approaches to ${\cal G}(\lambda,\gamma)$ from below.  Here, the sufficient condition for the discontinuous transition of the optimal lag time is ${\cal G}_I(0,\gamma)<0$ and $\partial{\cal G}_I(\lambda,\gamma)/\partial \lambda|_{\lambda=0}<0$. If $\tlb$ is greater than zero, the calculation gives the same value for arbitrary distribution functions $q(T)$ as shown below 
\begin{eqnarray*}
\lim_{\lam\to +0}{\cal G}_I&=&\lim_{\lam\to +0}\frac{\partial \bigl(\int_{\tlb}^\tub q(T)\ln f(T)dT\bigr)/\partial \lambda }{\partial \ln f(0)/\partial \lambda}\\
&=&\int_{\tlb}^\tub q(T)\lim_{\lam\to +0}\frac{\partial \ln f(T)/\partial \lambda}{\partial \ln f(0)/\partial \lambda}dT\\
&=&-\gamma
\end{eqnarray*}
\begin{eqnarray*}
\lim_{\lam\to +0}\frac{\partial}{\partial \lambda}{\cal G}_I&=&\lim_{\lam\to +0}\frac{\partial}{\partial \lambda} \frac{\partial \bigl(\int_{\tlb}^\tub q(T)\ln f(T)dT\bigr)/\partial \lambda }{\partial \ln f(0)/\partial \lambda}\\
&=&\int_{\tlb}^\tub q(T)\lim_{\lam\to +0}\frac{\partial}{\partial \lambda}\frac{\partial \ln f(T)/\partial \lambda}{\partial \ln f(0)/\partial \lambda}dT\\
&=&-\gamma(1+\gamma)
\end{eqnarray*}
Thus, the sufficient condition is fulfilled for any $q(T)$.\qed

Note that $T\to +0$ and $\lambda\to +0$ limit are non-communicative because 
$$\lim_{\lambda\to +0}\lim_{T\to +0}\frac{\partial^n f}{\partial \lambda^n}=(-1)^n\frac{(M+n)!}{M!}$$ 
and
$$\lim_{T\to  +0}\lim_{\lambda\to +0}\frac{\partial^n f}{\partial \lambda^n} =\gamma^n\frac{(M+n)!}{M!}$$ 
hold. Therefore, $\tlb>0$ is needed for the calculations from the fist line to the second line in above equations. However, since we cannot carry out the calculation with $\tlb=0$, the necessity of $\tlb>0$ is not proven. 

For the discrete distribution $\{q_i\}_{i=0}^{N-1}$ where $q_i$ represents the probability of the antibiotics application time to be $T_i$, we can use $\sum_{i=0}^{N-1}q_i\delta(T-T_i)$ as $q(T)$. Since the lower bound of the support of $q(T)$ is now always non-zero, the optimal lag time shows the discontinuous transition as $p$ increases. 

This condition is shown also as the necessary condition for the delta function-type model described in Section.\ref{sec:delta} where the role of a non-zero lower bound of $q(T)$ is explored more in detail. 

\section{Optimal lag time distributions}
\subsection{The general form of the optimal lag time distribution}
In this section, we show that the optimal lag time distribution (without specifying details of the waking-up dynamics) has the delta function at the origin, including the case that its coefficient is zero, and a gap region next to the origin in which the probability is zero regardless of the values of $p,\gamma$, and the shape of $q(T)$.

First of all, we point out that all the piecewise continuous probability distribution functions can be decomposed as
  $$r_\alpha(l)=\alpha\delta(l)+(1-\alpha)r_0(l)$$
with $\alpha\in[0,1]$, where $\delta(l)$ is the delta function peaked at the origin and $r_0(l)$ is a piecewise continuous probability distribution function satisfying $r_0(0)<\infty$. Note that $r_0(l)$ is also allowed to have at most a finite number of the delta functions. 

Here we assume that $r_\alpha(l)$ is the optimal lag time distribution with $r_0(l)$ which is chosen from the set of the probability distribution functions with zero lower bound of its support\footnote{We can construct such $r_\alpha(l)$ as following; first, find $r_0(l)$ from the set of the probability distribution function with zero lower bound of its support and without the delta function at the origin. Since $F[\alpha \delta(l)+(1-\alpha)r_0(l)]\geq F[r_0(l)]$ holds for any $r_0(l)$ including the case that $\alpha=0$ (the case with $\alpha>0$ will be shown later), we can choose optimal $\alpha$ for this $r_0(l)$.}. We suppose $\alpha<1$ in the following argument without losing generality because if $\alpha=1$ is the optimal value, we can regard the support of $r_0$ as $\emptyset$ and minsupp$(r_0)=\infty$. 

We add an extra argument $\tau$ indicating the lower bound of the support of $r_0$ to call $r_0(l,\tau)$ as the truncated distribution function while $r_0(l,0)$ is the original $r_0(l)$. One of the natural ways to define the truncated distribution function from $r_0(l)$ is 
\begin{eqnarray*}
r_0(l,\tau)=
\begin{cases}
0 & (l<\tau)\\
r_0(l,0)/\bigl(\int_\tau^\infty r_0(x,0)dx\bigr),& (l\geq \tau)
\end{cases}
\end{eqnarray*}
so that $r_0(l,\tau)$ is normalized. The support can consist of more than one disconnected intervals.

To see if having a non-zero lower bound increases the fitness value, we evaluate the derivative of the average fitness $F$  
\begin{eqnarray*}
    F[r,q](\gamma,p)&=&(1-p)\ln\Biggl[\int_{0}^{\infty}e^{-l}r(l)dl\Biggr]\\
    &+&p\int_0^\infty q(T)\ln\Biggl[e^{-(1+\gamma)T}\int_0^Te^{\gamma l}r(l)dl+\int_T^\infty e^{-l}r(l)dl\Biggr]dT.\nonumber
\end{eqnarray*}
with $r=r_\alpha(l,\tau)$ by $\tau$ at $\tau=0$. Because the single-round fitness $f(T)$ is given as 
\begin{equation}
    f(T)=
    \begin{cases}
    \alpha e^{-(1+\gamma)T}+(1-\alpha)\int_{\tau}^\infty e^{-x}\frac{r_0(x,0)}{\int_\tau^\infty r_0(y,0)dy}dx & (T<\tau)\\
    e^{-(1+\gamma)T}\Bigr(\alpha+(1-\alpha)\int_\tau^{T}e^{\gamma x}\frac{r_0(x,0)}{\int_\tau^\infty r_0(y,0)dy}dx\Bigr)+(1-\alpha)\int_{T}^\infty e^{-x}\frac{r_0(x,0)}{\int_\tau^\infty r_0(y,0)dy}dx & (T\geq \tau)
\end{cases}
\end{equation}
the derivative is calculated as 
\begin{eqnarray*}
\frac{\partial F}{\partial \tau}(\tau)&=&\int_0^\infty \hq(T)\frac{\partial \ln[f(T)]}{\partial \tau}dT\\
&=&\int_0^\tau \frac{\hq(T)}{f(T)}\frac{\partial }{\partial \tau}\Biggl[\alpha e^{-(1+\gamma)T}+(1-\alpha)\int_{\tau}^\infty e^{-x}\frac{r_0(x,0)}{\int_\tau^\infty r_0(y,0)dy}dx\Biggr]dT\\
&+&\int_\tau^\infty \frac{\hq(T)}{f(T)}\frac{\partial }{\partial \tau}\Biggl[ e^{-(1+\gamma)T}\Bigr(\alpha+(1-\alpha)\int_\tau^{T}e^{\gamma x}\frac{r_0(x,0)}{\int_\tau^\infty r_0(y,0)dy}dx\Bigr)\\
&+&(1-\alpha)\int_{T}^\infty e^{-x}\frac{r_0(x,0)}{\int_\tau^\infty r_0(y,0)dy}dx\Biggr]dT\\
&=&(1-\alpha)\int_0^\tau \frac{\hq(T)}{f(T)}\Biggl[-e^{\tau}\frac{r_0(\tau,0)}{\int_\tau^\infty r_0(y,0)dy}+\int_{\tau}^\infty e^{-x}\frac{\partial}{\partial \tau}\frac{r_0(x,0)}{\int_\tau^\infty r_0(y,0)dy}dx\Biggr]dT\\
&+&(1-\alpha)\int_\tau^\infty \frac{\hq(T)}{f(T)}\Biggl[-e^{-T-\gamma(T-\tau)}\frac{r_0(\tau,0)}{\int_\tau^\infty r_0(y,0)dy}+\int_\tau^{T}e^{\gamma x}\frac{\partial }{\partial \tau}\frac{r_0(x,0)}{\int_\tau^\infty r_0(y,0)dy}dx\\
&+&\int_{T}^\infty e^{-x}\frac{\partial }{\partial \tau}\frac{r_0(x,0)}{\int_\tau^\infty r_0(y,0)dy}dx\Biggr]dT\\
\end{eqnarray*}
with $\hq(T)$ as $(1-p)\delta(T)+pq(T)$.
From the definition of $r_0(x,\tau)$, $$\frac{\partial}{\partial\tau}r_0(x,\tau)=\frac{r_0(\tau,0)r_0(x,0)}{\Bigl(\int_\tau^\infty r_0(y,0)dy\Bigr)^2}\to r_0(0,0)r_0(x,0),\ ({\rm as\ }\tau\to 0)$$
holds. By substituting $\tau=0$ to the equation above, we get
\begin{eqnarray}
c\frac{\partial F}{\partial \tau}\Biggr|_{\tau=0}
&=&\int_0^\infty \frac{\hq(T)}{f(T)}\Biggl(e^{-(1+\gamma)T}\int_0^Te^{\gamma x}r_0(x,0)dx+\int_T^\infty e^{-x}r_0(x,0)dx\Biggr)dT\nonumber\\
&-&\int_0^\infty \frac{\hq(T)}{f(T)}e^{-(1+\gamma)T}dT\label{eq:dfdtau}
\end{eqnarray}
where $c = ((1-\alpha)r_0(0,0))^{-1}$ ($0<c<\infty$). To judge whether a non-zero lower bound of the support is advantageous, we want to see the size relationship between the first and the second term in the right hand side of the equation.

Now we calculate the functional variation of $F$ by $r_\alpha$. Consider the situation that $r_\alpha(x)$ is perturbed by an arbitrary probability distribution function with zero lower bound of the support, $\eta(x)$, as $r_\alpha(x)\rightarrow r_\alpha'(x)\propto r_\alpha(x) + \epsilon \eta(x)$. Since $r_\alpha'(x)$ still needs to be a probability distribution function, it must be normalized and $\epsilon$ has to be positive. Since we chose $\eta(x)$ from the probability distribution functions, $r_\alpha'(x)$ is given as $ (r_\alpha(x) + \epsilon \eta(x))/(1+\epsilon)$. By taking the derivative of $F$ with $r_\alpha'(x)$ by $\epsilon$ at $\epsilon=0$, we get
\begin{eqnarray}
\frac{\partial F}{\partial \epsilon}\Biggr|_{\epsilon=0}&=&\int_0^\infty \frac{\hq(T)}{f(T)}\Biggl(e^{-(1+\gamma)T}\int_0^Te^{\gamma x}(\eta(x)-r_\alpha(x))dx\nonumber \\&+&\int_T^\infty e^{-x}(\eta(x)-r_\alpha(x))dx\Biggr)dT\label{eq:variation}
\end{eqnarray}
Note that $f(T)$ is calculated with $r_\alpha(x)$, but not $r_\alpha'(x)$ because $\epsilon=0$ is substituted. Because of the optimality, $\partial F/\partial \epsilon$ at $\epsilon=0$ needs to be smaller than or equals to zero. This is not necessarily the extreme point as usual functional variations lead because the perturbation with $\epsilon<0$ is not allowed in this problem. However, if it is zero, the functional $I$ defined by
$$I[a](T)=e^{-(1+\gamma)T}\int_0^Te^{\gamma x}a(x)dx+\int_T^\infty e^{-x}a(x)dx$$with $a$ as a function with zero lower bound of its support needs to satisfy the equality 
$$I[\eta_0](T)=I[r](T)=I[\eta_1](T)$$
for ${}^\forall T\geq0$ where $\eta_0$ and $\eta_1$ are different probability distribution function because $q(T)$ is an arbitrary function. By choosing, for instance, an exponential distribution with the parameter $\lambda_i$ ($\lambda_0\neq \lambda_1$) as $\eta_i(x)$, we can easily show that $I[\eta_0]\neq I[\eta_1]$. Thus, $\partial F/\partial \epsilon$ needs to be smaller than zero.

By choosing the delta function $\delta(x)$ as $\eta(x)$ the optimal condition (right hand side of Eq.(\ref{eq:variation}) is smaller than zero) leads to 
$$\int_0^\infty \frac{\hq(T)}{f(T)}\Biggl(e^{-(1+\gamma)T}\int_0^Te^{\gamma x}r_0(x,0)dx+\int_T^\infty e^{-x}r_0(x,0)dx\Biggr)dT > \int_0^\infty \frac{\hq(T)}{f(T)}e^{-(1+\gamma)T}dT.$$
This is  what we desired. From Eq.(\ref{eq:dfdtau}) and the inequality above, $\partial F/\partial \tau\bigr|_{\tau=0}$  is shown to be positive, and thus, the non-zero lower bound increases the fitness. (Note that it is possible that $r_0(l,\tau)$ is suboptimal among all the function with its lower bound of the support as $\tau$, while still it performs better than any distribution function with zero lower bound of the support.). 

Therefore, if $r_\alpha$ is the optimal distribution, there is always a region next to the origin in which $r_\alpha$ is zero. \qed

So far, we have assumed that $0<\alpha<1$ can be optimal, but the existence of the optimal $\alpha$ in $(0,1)$ is not yet shown. The remaining part of this subsection is devoted for showing it.

We evaluate the first-order derivative of $F[r_\alpha]$ by $\alpha$ at $\alpha=0$, given as
\begin{eqnarray*}
\frac{\partial F}{\partial \alpha}&=&\int_0^\infty \hq(T)\frac{\partial}{\partial \alpha}\ln[f(T,\alpha)]dT\\
\end{eqnarray*}
where we added $\alpha$ as an argument of $f$ to emphasize that $f$ is a function of $\alpha$. The existence of $\alpha$ satisfying of $\partial F/\partial \alpha=0$ in $[0,1]$ is equivalent to that the optimal $\alpha$ is in $[0,1]$, and otherwise, the optimal $\alpha$ is either zero or unity. Therefore, if we show the uniqueness of the solution in $[0,1]$ for given $(\gamma,p,q(T))$, the continuity of the optimal $\alpha$ on $(\gamma,p,q(T))$ will be shown because the derivative is a continuous function of $\gamma,p$, and $q(T)$ (for $q(T)$, we need to calculate the functional variation of $F$). By noting that $f$ is linear in $\alpha$, $\partial_\alpha f$ is a constant function in $\alpha$. Therefore, the second order derivative is given as
\begin{eqnarray*}
\frac{\partial^2 F}{\partial \alpha^2}&=&-\int_0^\infty \hq(T)\frac{(\partial_\alpha f)^2}{f^2}dT<0
\end{eqnarray*}
meaning that $\partial_\alpha F$ is a monotonically decreasing function of $\alpha$, and thus, there is at most one solution in $[0,1]$. Now the optimal $\alpha$, $\alpha^*=\alpha^*(\gamma,p)[q]$ is a continuous function. 

$\alpha^*(\gamma,0)[q]=1$ holds because of the monotonicity  of $e^{-x}$. Also, there is the critical values of $p$ and $\gamma$ at which the optimal $\alpha$ transits to zero if the delta function is chosen as $q(T)$ (see Section.\ref{sec:delta}). Thus, there is a way to change the optimal $\alpha$ continuously from unity to an arbitrary value by continuously modulating $(p,\gamma,q(T))$.

\subsection{Computing the optimal lag time distribution}

\subsubsection{Derivation of the optimal condition}
In this section, we present the method to obtain the optimal lag time distribution for a given distribution of the antibiotics application time, $q(T)$. If the distribution is discrete, given as ${\rm Prob}(T_i)=q_i,\ (T_i>0)$ for $i=1,\cdots,M-1$ and $\sum_{i=1}^{M-1}q_i=1$, we introduce the expression $q(T)=\sum_{i=1}^{M-1}q_i\delta(T-T_i)$ for the unified description.

Here, we compute the optimal lag time distribution for $q(T)$ with the finite support $[0,T_{\rm max})$ while the case for the infinite support can be considered by taking $T_{\rm max}\to\infty$ limit \footnote{Note that the choice of $T_{\rm max}$ corresponds to selecting a way of the antibiotics application. Even if two distribution functions $q_1(T)$ and $q_2(T)$ with the support $[0,T^{(1)}_{\rm max})$ and $[0,T^{(2)}_{\rm max})$, respectively, converge to the same distribution $q(T)$ in $T^{(1)}_{\rm max},T^{(2)}_{\rm max}\to \infty$ limit, the optimal lag time distributions for $q_1(T)$ and $q_2(T)$ are different in general.}. Then, the upper bound of the optimal lag time distribution becomes $T_{\rm max}$ because it just decreases the fitness to invest any fraction of population to have lag times longer than $T_{\rm max}$. (For the effect of $T_{\rm max}$ to the optimal lag time, see Fig.S\ref{fig:Tmax}).

Now the fitness function is given as 
$$F=\int_0^{T_{\rm max}}\hq(T)\ln\Bigl[\int_{0}^{T_{\rm max}}h(T,l)r(l)dl\Bigr]dT$$
where $\hat{q}(T)=(1-p)\delta(T)+pq(T)$ and $h(T,l)$ is
\begin{equation}
h(T,l)=
\begin{cases}
\exp[-T-\gamma(T-l)]\ &(l<T)\\
\exp[-l]\ &({\rm otherwise}).
\end{cases}
\end{equation}
Next, we approximate the optimal lag time distribution by discretizing it with the bins of size $\Delta$, $\{r_n\}_{n=0}^{N-1}$ where $N=T_{\rm max}/\Delta$. Now we obtain
$$F_d(T)=\int_0^{T_{\rm max}}\hq(T)\ln\Bigl[\sum_{n=0}^{N-1}h(T,n\Delta )r_n\Bigr]dT.$$
 Owing to the discretization, now we can carry out the maximization of the function $F_d$ by taking differentials of $F_d$ respect to $r_n'$s. Since $r_n'$s are the discrete probability distribution function, $r_n\geq 0$ and $\sum_{n=0}^{N-1}r_n=1$ needs to be satisfied.

 This can be solved by introducing KKT multipliers $\mu_n$ for the condition $r_n\ge 0$ and $\bar \mu$ for the condition $\sum_{n=0}^{N-1}r_n=1$ as 
 $$
 L(\vec{r},\vec{\mu})=\int_0^{T_{\rm max}}\hq(T)\ln\Biggl[\sum_{n=0}^{N-1}h(T,n\Delta )r_n \Biggr]dT+\sum_{n=0}^{N-1}\mu_n r_n-\bar \mu \sum_{n=0}^{N-1}r_n,
 $$
 and imposing 
 \[\frac{\partial L}{\partial r_n}=0,\ \ \mu_nr_n=0,\ \mu_n\geq 0,\ r_n\geq 0,\ (0\leq n<N), \sum_{n=0}^{N-1}r_n=1.
 \]
This results in 
\begin{eqnarray}
\bar \mu-\mu_n&=&\int_0^{T_{\rm max}}\hq(T)\frac{h(T,n\Delta)}{\langle h(T)\rangle}dT \label{seq:condition1}\\
r_n=0&{\rm \ or}&\ \mu_n=0.
\end{eqnarray}
with $r_n\geq 0$ and $\mu_n\geq 0$ for $n=0,1,\cdots,N-1$. $\langle h(T)\rangle$ represents the mean, i.e., $\langle h(T)\rangle=\sum_{n=0}^{N-1}h(T,n\Delta)r_n$. 
The Lagrange multiplier $\bar\mu$ is determined by 
multiplying eq.~(\ref{seq:condition1}) with $r_n$ and taking the sum over $n$. 
Knowing $r_n\mu_n=0$, this leads to 
\[
\bar \mu\left(\sum_{n=0}^{N-1} r_n\right)=\int_0^{T_{\rm max}}\hq(T)\frac{\sum_{n=0}^{N-1}r_n h(T,n\Delta)}{\langle h(T)\rangle}dT 
=\int_0^{T_{\rm max}}\hq(T)dT=1, 
\]
giving $\bar \mu=1$.

Next, we carry out the integral over $T$. If the distribution of $T$ is discrete, the integral leads to 
$$\int_0^{T_{\rm max}}\hq(T)\frac{h(T,n\Delta)}{\langle h(T)\rangle}dT=\sum_{m=0}^{M-1}\hq_m\frac{h(T_m,n\Delta)}{\langle h(T_m)\rangle},$$
where $\hq_0=(1-p)$ and $\hq_m=pq_m\ (m\geq 1)$. 

If $q(T)$ is a continuous function, since $h(T,l)$ has different expression depending on the size relationship between $T$ and $l$, we divide the integral into parts so that $h(T,l)$ has the same expression in each integral as follows;
\begin{equation}
\int_0^{T_{\rm max}}\hq(T)\frac{h(T,n\Delta)}{\langle h(T)\rangle}dT=\sum_{m=0}^{N-1}\int_{m\Delta}^{(m+1)\Delta}\hq(T)\frac{h(T,n\Delta)}{\langle h(T)\rangle}dT.
\end{equation}

Then, we assume that $\Delta$ is sufficiently small so that each integral is well-approximated by
$$\hq_m\frac{h(m\Delta,n\Delta)}{\langle h(m\Delta)\rangle},$$
where $\hq_m$ is $\Delta\cdot \hq(m\Delta)$ for $m\neq 0$ while $\hq_0$ by $(1-p)+pq(0)\Delta$. Now, by rewriting $h(m\Delta,n\Delta)$ and $\langle h(m\Delta)\rangle$ (or $h(T_m,n\Delta)$ and $\langle h(T_m)\rangle$ in the discrete case) as $h^m_n$ and $\langle h^m\rangle$, the condition for the $n$th bin is given as

\begin{equation}
1-\mu_n=\sum_{m=0}^{S-1}\hq_m\frac{h^m_n}{\langle h^m\rangle},\ (r_n=0{\rm \ or}\ \mu_n=0)\label{eq:lagrange_cont_sum}
\end{equation}
with $r_n\geq 0$ and $\mu_n\geq 0$, where $S=N$ for a continuous $q(T)$ while $M$ for discrete $q(T)$. 

A remarkable feature of the equations is that the distribution  $\{r_i\}_{i=0}^{N-1}$ appears only in the form of the averages and the equations are independent linear equations in $1/\langle h^{m}\rangle$. Because of the linearity of the equations, as many equalities as the number of terms in the sum are satisfied. If $q(T)$ is the continuous function, the number of non-zero $\hq_m'$s equals to $N$, and thus, all $N$ equations are solvable while it is not guaranteed that the solution satisfies the constraint $r_i\geq0$. In contrast, if $q(T)$ is a discrete function, the number of averages appearing in the sum can be either less or more than $N$. However, if we take $\Delta$ sufficiently small, $M<N$ generally holds. Hereafter, we suppose $q$ is a continuous distribution function, and accordingly, the number of non-zero $\hq_m'$s is $N$. 

It is proven in the previous section that the optimal lag time distribution generally has a gap region next to the origin. Thus, some $r_n'$s need to be set to zero meaning that we cannot use the equation (Eq.(\ref{eq:lagrange_cont_sum})) for some $n'$s to determine the optimal lag time distribution. However, it is worth mentioning that if we virtually remove the constraints $r_n\geq 0$ (then $\mu_n=0$ is the solution for all $\mu_n'$s), we can solve $r_n'$s via solving the average $\langle h^m\rangle$. 

By denoting $1/\langle h^m\rangle$ by $x_m$ and supposing $x_m's$ as variables, Eq.(\ref{eq:lagrange_cont_sum}) is regarded as a system of linear equations, and thus, the solution is given as ${\bf x}=A^{-1}{\bf 1}$ where
\[A=
  \left(
    \begin{array}{cccc}
      \hq_0h_0^{0} & \hq_1h_0^{1} & \ldots & \hq_{N-1}h_0^{{N-1}} \\
     \hq_0h_1^{0} & \hq_1h_1^{1} & \ldots & \hq_{N-1}h_1^{{N-1}}  \\
      \vdots & \vdots & \ddots & \vdots \\
     \hq_0h_{N-1}^{0} & \hq_1h_{N-1}^{1} & \ldots & \hq_{N-1}h_{N-1}^{{N-1}} 
    \end{array}
  \right),\ \  
  {\bf 1}=
  \left(
  \begin{array}{c}
      1\\
      1 \\
      \vdots \\
      1
      \end{array}
  \right).
\] 
From the definition of the average, the lag time distribution $\{r_i\}_{i=0}^{N-1}$ are given as ${\bf r}=B^{-1}{\bf x}_{\rm inv}$, where
\[B=
  \left(
    \begin{array}{cccc}
      h_0^{0} & h_1^{0} & \ldots & h_{N-1}^{0} \\
     h_0^{1} & h_1^{1} & \ldots & h_{N-1}^{1} \\
      \vdots & \vdots & \ddots & \vdots \\
  h_0^{{N-1}} & h_1^{{N-1}} & \ldots & h_{N-1}^{{N-1}}
    \end{array}
  \right)
   ,\ \ {\bf x}_{\rm inv}=
  \left(
  \begin{array}{c}
      1/x_0\\
      1/x_1 \\
      \vdots \\
      1/x_{N-1}
      \end{array}
  \right).
\]

\subsubsection{Solving the equations}
As discussed, some of the conditions are not fulfilled without setting $r_i$ to zero. Therefore, the optimal lag time distribution is never obtained by solving the linear equation above. Instead, we need to set some $r_i$'s to zero and solve the equations. Since the $i$th equation,
$$1-\mu_i=\sum_{j=0}^{N-1}\hq_j\frac{h^j_i}{\langle h^j\rangle}$$
is no longer the condition to determine the optimal distribution then, the system equations to determine the average is now underdetermined. Thus, we first solve the average and the probability as the function of  undetermined averages, and consequently, solve the self-consistent equations for the undetermied averages.

We re-order indices so that $r_i'$s with $i\in\{L\cdots N-1\}\equiv K$ are set to zero, and leave $x_i$ with $i\in K$ as free parameters to solve $x_i$ with $i\in \{0,1,\cdots,L-1\}$ (There is no criterion to find the index set before solving the equations). We refer the $L$ by $L$ submatrix of $A$ by $A_L$.  
Now, $N$ linear equation with $N$ variable becomes $L$ equations with $L$ variable given as

\[
  \left(
    \begin{array}{cccc}
      a_{00} & a_{01} & \ldots & a_{0(L-1)} \\
  a_{10} & a_{11} & \ldots & a_{1(L-1)} \\
      \vdots & \vdots & \ddots & \vdots \\
      a_{(L-1)0} & a_{(L-1)1} & \ldots & a_{(L-1)(L-1)} \\
    \end{array}
  \right)
  \left(
  \begin{array}{c}
      x_0\\
      x_1 \\
      \vdots \\
      x_{L-1}
      \end{array}
  \right)
  =
  \left(
  \begin{array}{c}
      1-\sum_{i=L}^{N-1} a_{0i}x_i\\
      1-\sum_{i=L}^{N-1} a_{1i}x_i \\
      \vdots \\
      1-\sum_{i=L}^{N-1} a_{(L-1)i}x_i
      \end{array}
  \right)
\]
where $a_{ij}$ is the $(i,j)$th element of the original $N$ by $N$ matrix $A$ after re-ordering indices. The solutions are given as $$x_n=\Xi_n-\sum_{j=L}^{N-1}\xi_{ni}x_i,\ (0\leq n<L),$$
where $\Xi_n=\sum_{i=0}^{L-1}(A_L^{-1})_{ni}$ and $\xi_{ni}=\sum_{j=0}^{L-1}(A_L^{-1})_{nj}a_{ji}$. Similarly, we use the submatrix of $B$, $B_L$ to solve
\[
  \left(
    \begin{array}{cccc}
      b_{00} & b_{01} & \ldots & b_{0(L-1)} \\
  b_{10} & b_{11} & \ldots & b_{1(L-1)} \\
      \vdots & \vdots & \ddots & \vdots \\
      b_{(L-1)0} & b_{(L-1)1} & \ldots & b_{(L-1)(L-1)} \\
    \end{array}
  \right)
  \left(
  \begin{array}{c}
      r_0\\
      r_1 \\
      \vdots \\
      r_{L-1}
      \end{array}
  \right)
  =
  \left(
  \begin{array}{c}
      1/x_0\\
     1/x_1\\
      \vdots \\
      1/x_{L-1}
      \end{array}
  \right).
\]Now $r_i's (0\leq i<L)$ are solved as functions of undetermined averages, $x_L,x_{L+1},\cdots,x_{N-1}$. As the last step, we need to consistently determine the averages $x_L,x_{L+1},\cdots,x_{N-1}$. The equations is given as

\[
  \left(
    \begin{array}{cccc}
      b_{L0} & b_{L1} & \ldots & b_{L(L-1)} \\
  b_{(L+1)0} & b_{(L+1)1} & \ldots & b_{(L+1)(L-1)} \\
      \vdots & \vdots & \ddots & \vdots \\
      b_{(N-1)0} & b_{(N-1)1} & \ldots & b_{(N-1)(L-1)} \\
    \end{array}
  \right)
  \left(
  \begin{array}{c}
      r_0(x_L,x_{L+1},\cdots,x_{N-1})\\
      r_1(x_L,x_{L+1},\cdots,x_{N-1}) \\
      \vdots \\
      r_{L-1}(x_L,x_{L+1},\cdots,x_{N-1})
      \end{array}
  \right)
  =
  \left(
  \begin{array}{c}
      1/x_L\\
     1/x_{L+1}\\
      \vdots \\
      1/x_{N-1}
      \end{array}
  \right).
\]
Note that here the matrix in the left hand side of the equation is $(N-L)$ by $L$, but not square matrix. Therefore, we cannot invert the matrix, but instead, derive the self-consistency equations as follows;
\begin{equation}
x_{L+n}\sum_{i=0}^{L-1}\frac{\tilde{\eta}_{ni}}{1-\sum_{j=L}^{N-1}\tilde{\xi}_{ij}x_j}-1=0 \label{eq:z_Lth}
\end{equation}
with $\tilde{\xi}_{ij}=\Xi_i^{-1}\xi_{ij}$ and $\tilde{\eta}_{ij}=\Xi_j^{-1}\sum_{k=0}^{L-1}b_{(i+L)k}(B_L^{-1})_{kj}$. By solving the system of $L$th order equation, we obtain the solutions of $x_i's$ with $i\in K$ and accordingly remaining $x_i's$ and the lag time distribution $\{r_i\}_{i=0}^{N-1}$.  While Eq.(\ref{eq:z_Lth}) is not solvable symbolically, the solution of ${\bf x}$ obtained by solving the linear equations above works well as the initial guess of numerical solvers.

\subsubsection{Computational protocols}
There is no criterion to find the indices which cannot satisfy the conditions without setting $r_i=0$, and thus, in principle, only the way to obtain the optimal distribution $\{r_i\}_{i=0}^{N-1}$ is trying all the possible combination of the indices that $r_i$ is set to zero, check if all $r_j'$s which are not initially set to zero are positive, and compare the fitness of all those consistent solutions.

It is practically impossible. Thus, we take advantage of the proven form of the optimal lag time distribution. Recall that the optimal lag time has a form $u(l)=\alpha\delta(l)+(1-\alpha)u_0(l)$ with minsupp$(u_0)>0$. Thus, we need to know how many connected components configure the support of $u_0(l)$. If the number of connected components is just one, we can carry out the computation to check all the possible zero/non-zero allocation for $r_i'$s. While this is nothing but a heuristic, we have performed numerical optimizations of the fitness function by $r_i$ values to guess the number of the connected components. The optimization indicated that the number of the connected components is the same number with the peaks of $q(T)$. Therefore, we computed all the possible distributions with a single connected support of $r_i$ ($i\geq 2$)\footnote{We checked only for $i\geq 2$ because $i=0$ is for the delta function and at least $i=1$ is for a gap as long as the discretization is done well to approximate the distributions}. The optimal distributions for a normal, exponential, and power-law distribution are computed in this way. This method guarantees that the obtained distribution has the highest fitness among all the locally optimal distributions whose support has two elements (origin and the support of $u_0(l)$).

Still, too much intensive computation is required if the number of connected components of supp$(u_0)$ is more than one. Thus, for the computation of the optimal lag time distribution for a double-Gaussian distribution shown in the main text, we adopted an iterative solving technique. The scheme is following:
\begin{enumerate}
    \item[(0)] solve the linear equation to obtain ${\bf r}=B^{-1}{\bf x}_{\rm inv}$ add the indices $n$ to $K$ if $r_n$ becomes smaller than or equal to zero in the solution.
    \item[(1)] solve the consistency equation 
    $$x_{L+m}\sum_{i=0}^{L-1}\frac{\tilde{\eta}_{mi}}{1-\sum_{j=L}^{N-1}\tilde{\xi}_{ij}x_j}-1=0$$
    to obtain $r_n'$s, where $L+m\in K\ (m=0,1,\cdots\# K-1)$.
    \item[(2)] if all the bin $r_n'$s are greater than or equal to zero, then end the iteration, otherwise, put the indices $n$ whose $r_n$ are negative to $K$ and return to (1). 
\end{enumerate}
The definition of $K$ is given in the second paragraph of this section, and at every update of $K$, the indices are re-ordered so that $K$ is written as $\{L,\cdots,N-1\}$. It is possible that the obtained distribution by this method is inferior to other distributions even if those distributions have the same or less numbers of the elements of the support. However, it is confirmed that this iterative method gives the same distribution led by the former method for normal and exponential distributions, but not power-law distributions as far as we have tried.

\begin{figure}[htbp]
\begin{center}
\includegraphics[width = 130 mm, angle = 0]{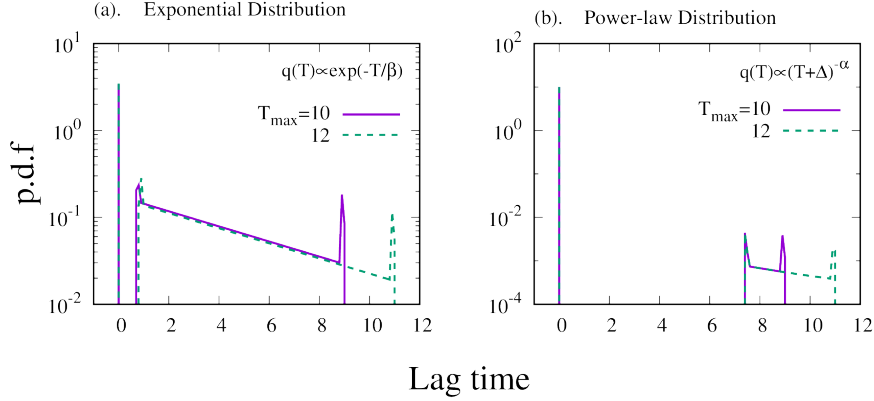}
\caption{The optimal distribution is computed for (a). the exponential and (b) the power-law distribution, respectively, with two different values of $T_{\rm max}$. Here, we compare the optimal distributions with $T_{\rm max}=10$ and $12$. While the exponential distribution and the power-law distribution led to the distinct optimal distributions for the different $T_{\rm max}$ values, the optimal distributions for the normal distribution and the sum of the two normal distributions (panel (a) and (d) in Fig.5 in the main text) were unchanged. The other parameter values are the same with Fig.5 in the main text.}
    \label{fig:Tmax}
  \end{center}
\end{figure} 

\section{The delta function-type model}
\label{sec:delta}
We can also consider the simplest model where the cells start to grow immediately after the time $\lam$ has passed without any variation, i.e., the lag time distribution follows the Dirac's delta function. Since the model allows us to extract the mathematically clear essences of the population dynamics of waking-up, in this section we repeat the same analysis done for the sequential models. Some results obtained from the delta function-type model differ from the outcome of the sequential model even qualitatively. Still, it provides useful insights to us for understanding the results described above and in main text. 

\subsection{The optimal lag time in a single, and double phenotypes case}
Therefore, the temporal evolution of the bacterial population at time $t \ (t>{\rm max}\{\lam,T\})$ given as following (Table.\ref{table:tab1}),
\begin{table}[htb]
\begin{center}
  \begin{tabular}{|c|c|c|} \hline
    & + AB  (prob. $p$) &- AB (prob. $1-p$)\\ \hline
    $\lam<T$ & $\exp[-\gamma(T-\lam)]\exp[t-T]$ & $\exp[t-\lam]$\\ \hline
    $\lam>T$ & $\exp[t-\lam]$ & $\exp[t-\lam]$\\ \hline
  \end{tabular}
  \caption{The population of the cells at time $t>\rm{max}\{\lam,T\}$ for each condition.}
  \label{table:tab1}
\end{center}
\end{table}

Then, the fitness function is given by
\begin{equation}
F_I^\delta(\lam,\gamma,p,T) = \begin{cases}
-p(T-\lam)(1+\gamma)-\lam & (\lam<T)\\
-\lam &(\lam>T).
\end{cases}\label{eq:fit_delta}
\end{equation}
By taking the first-order derivative of $F_I^\delta$ respect to $\lam$, we obtain 
\begin{equation}
\frac{\partial F_I^\delta}{\partial \lam}(\lam,\gamma,p,T) = \begin{cases}
p(1+\gamma)-1 & (\lam<T)\\
-1 &(\lam>T).
\end{cases}\label{eq:derivative_delta}
\end{equation}
Therefore, the optimal lag time, $\lam^*$ shows the discontinuous transition as
\begin{equation}
    \lam^*=\begin{cases}
    0 & (\gamma < 1/p-1)\\
     T & (\gamma \geq1/p-1).\\ 
    \end{cases}
\end{equation}
It is worth noting that the transition point is determined only by $\gamma$ and $p$. In contrast to the models described above and in the main text. The antibiotics application time $T$ has no role in the transition. Also, the $\gamma=1/p-1$ is not upper bound of the transition point but exactly determines the transition point. 

The calculations for the two-strategy case are also carried out analytically. For the two species case, the optimal parameter values are given as $\lambda_a=0$, $\lambda_b=T$, and 
$$x^*=\frac{p(1-\exp[-(1+\gamma)T])-\exp[-\gamma T](1-\exp[-T]))}{(1-\exp[-T])(1-\exp[-\gamma T])}.$$
The optimal $x$ is shown in Fig.\ref{fig:delta_heatmap} which has qualitatively the same features of the optimal fraction of the model described in the main text. 

\begin{figure}[htbp]
\begin{center}
\includegraphics[width = 120 mm, angle = 0]{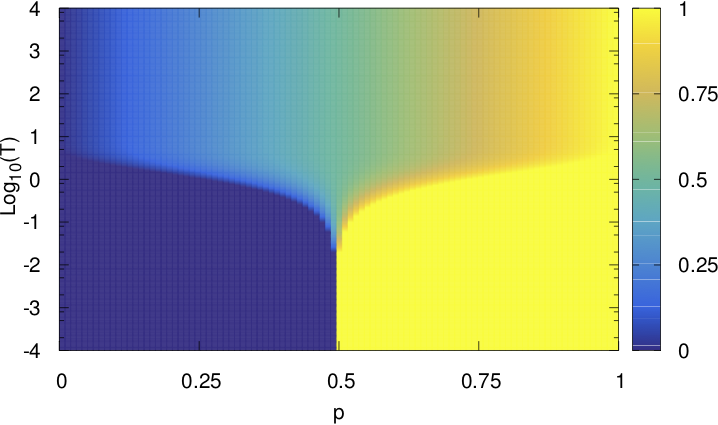}
\caption{The the optimal fraction $x^*$ is plotted as a function of $p$ and $T$. The steep transition at $p\approx 0.5$ in smaller $T$ region is continuous. $\gamma$ is set to unity. }
    \label{fig:delta_heatmap}
  \end{center}
\end{figure}

\subsection{The steepness of the transition and the support of $q(T)$}
We have shown that a non-zero lower bound of the support of $q(T)$ is a sufficient condition for the sequential model to exhibit the discontinuous transition of the optimal lag time. To see how the lower bound of the support changes nature of the transition, here we explicitly calculate the first order derivative of the fitness by the lag time $\lambda$. 

Here, a normal distribution with the average $\tau$ and the standard deviation $\sigma$ is chosen as the distribution function of $T$. Then, the fitness function is given in the form ${\cal F}_I^\delta=-(1-p)\lam+p{\cal G}_I^\delta(\lambda,\gamma)$ and ${\cal G}_I^\delta$ has three different expressions depending on the size relationship among $\tlb, \tub$, and $\lambda$ which are given as
\begin{eqnarray*}
{\cal G}_I^{\delta}=
\begin{cases}
\int_\tlb^\tub (\gamma\lambda-(1+\gamma)T)q(T)dT & (\lambda\leq\tlb)\\
-\int_\tlb^\lambda \lambda q(T)dT+\int_\lambda^\tub (\gamma\lambda-(1+\gamma)T)q(T)dT & (\tlb<\lam\leq\tub)\\ 
-\int_\tlb^\tub \lambda q(T)dT & (\tub<\lam).
\end{cases}
\end{eqnarray*}
In the third case, the integral simply leads to $-\lambda$ which means $\partial_\lam {\cal F}_I^\delta=-1$. Thus $\lambda>\tub$ never be an optimal solution and there is no need to consider this region. 

For the first case, by carrying out the integral, we obtain ${\cal G}_I^{\delta}=\gamma\lambda-{\cal H}_I^\delta(\gamma)$ where ${\cal H}_I^\delta$ is constant in $\lam$\footnote{It is given as $${\cal H}_I^\delta=\frac{\sqrt{2}\sigma(1+\gamma)}{\Omega}\Bigl((e^{-(\tau'-\tlb')^2}-e^{-(\tau'-\tub')^2})/\sqrt{\pi}+\tau'(\erf(\tau'-\tlb')-\erf(\tau'-\tub'))\Bigr),$$ where $\tau',\tlb',\tub'$, and $\Omega$ is $\tau/(\sqrt{2}\sigma)$, $\tlb/(\sqrt{2}\sigma)$, $\tub/(\sqrt{2}\sigma)$, and $\erf(\tau'-\tlb')-\erf(\tau'-\tub')$, respectively.}.  Therefore, the first-order derivative of ${\cal F}_I^\delta$ is given as 
$$\frac{\partial {\cal F}_I^\delta}{\partial \lam}=-(1-p)+p\gamma.$$
The derivative shows that the optimal lag time $\lambda^*$ is zero as long as $\gamma < (1-p)/p$ holds and it transits to a some value greater than $\tlb$. 

To see what happens if the lower bound of the support is zero, here we set $\tlb=0$. Also, for the simplicity, we assume $\tub\to\infty$. This assumption has no effect to the result because the derivative of the fitness function is always minus one in the third region $\lambda>\tub$. 

Since now the support is ${\mathbf R}^+$, ${\cal F}_I^\delta$ has only one expression that  
\begin{eqnarray*}
&&\frac{\Omega}{\sqrt{2}\sigma p}{\cal F}=-\lp\frac{1-p}{p}(1+\erf(\tp))\\&&+\lp(\gamma-\erf(\tp))-(1+\gamma)\Bigl[(\lp-\tp)\erf(\lp-\tp)+\tp+\frac{e^{-(\lp-\tp)^2}}{\sqrt{\pi}}\Bigr],
\end{eqnarray*}
where the variables with ${}'$ represent the original variables divided by $\sqrt{2}\sigma$ and $\Omega=1+\erf(\tp)$. Then, the first-order derivative by $\lam$ (note that it is not by $\lp$) is 
\begin{equation}
\frac{\Omega}{p(1+\gamma)}\frac{\partial {\cal F}}{\partial \lam}=- \erf(\lp-\tp)+h(\tp,\gamma,p)\label{eq:dF_delta_Normal}
\end{equation}
where $h$ is given as
$$h(\tp,\gamma,p)=\frac{1}{1+\gamma}\Bigl(\gamma-\erf(\tp)-\frac{1-p}{p}(1+\erf(\tp))\Bigr).$$
Let us consider the transition triggered by an increase of $p$. For the simplicity, we assume $\tp\gg1$ and $\gamma=1$ leading to $h\approx -(1-p)/p$. Since $h$ diverges to $-\infty$ as $p\to0$, the optimal lag time is zero at $p\approx 0$. As $p$ increases, $h$ approaches to $\erf(\lp-\tp)$ from below. By noting that the error function $\erf$ is the monotonically increasing function of the argument, there is only one intersection of $\erf(\lp-\tp)$ and $h$, and it is first made at $\lp=0$. Therefore, the transition is continuous. 

Note that $\lp$ and $\tp$ are now scaled by the standard deviation of Normal distribution, $\sigma$, and thus, the flatness of $\erf(\lp-\tp)$ in $\lp\ll\tp$ (and also $\lp\gg\tp$) and the steepness of its sigmoidal shape depends on $\sigma$. If the distribution is greatly broad ($\sigma \gg \tau$), the error function is approximately a linear function even around $\lambda=0$. With such broad distribution, the transition looks continuous. 

However, as $\sigma$ becomes smaller and smaller, the error function $\erf(\lp-\tp)$ around $\lambda=0$ becomes parallel to $\lambda$ axis, being much closer to the constant function. Then, the cross point of $\erf(\lp-\tp)$ and $h$ gets highly sensitive to small changes of $p$ value, looks more and more similar to the discontinuous transition. Especially, under the $\sigma\to 0$ limit, the error function converges to the step function $\Theta(\lambda-\tau)$ that the intersection with $h$ is possible only at $\lambda=\tau$ except $p=0.5$ at which $(1-p)/p$ completely overlaps to the step function in the region $\lambda<\tau$. Note that now the antibiotics application time is $\tau$ with no fluctuation, and thus, the transition of the optimal lag time from $0$ to $\tau$ corresponds to the discontinuous transition discussed in the previous section. 

\subsection{A sufficient and necessary condition for the discontinuous transition}
We showed that if and only if the support of Normal distribution has non-zero lower bound, the optimal lag time exhibits a discontinuous transition. Indeed, exactly the same argument is applied for an arbitrary probability distribution functions which has a well-defined mean. 

Suppose any distribution function $q(T)$ with its support $(T_{\rm lb},T_{\rm ub})$ where $T_{\rm ub}$ can be either finite or infinite. Here we assume that $q(T)$ has a well-defined mean.  The first-order derivative of the fitness function ${\cal F}_I^\delta$ respect to $\lambda$ is given as 
\begin{eqnarray*}
\frac{\partial{\cal F}_I^\delta}{\partial \lambda} &=&
  \begin{cases}
    -(1-p)+\gamma p & (\lambda<T_{\rm lb})\\
   p(1+\gamma)\Bigl( -{\cal G}_I^\delta(\lambda)+{\cal H}_I^\delta(\gamma,p)\Bigr)& (T_{\rm lb}\leq \lambda < T_{\rm ub})\\
    -1 &(T_{\rm ub}\leq \lambda),
  \end{cases}\\
  {\cal G}_{I}^\delta&=&Q(\lam)\\
  {\cal H}_I^\delta&=&\frac{1}{1+\gamma}\Bigr(Q(\tlb)+\gamma  Q(\tub)-\frac{1-p}{p}\Bigl)
\end{eqnarray*}
where $Q$ is an arbitrary chosen primitive function of the distribution function of $q$. Again, the optimal lag time is zero as long as $-(1-p)+\gamma p<0$ holds, and otherwise, it transits discontinuously to a non-zero value determined by the second case of above equation. 

To trigger the transition of the optimal lag time from zero to non-zero, ${\cal H}_I^\delta$ must approach to ${\cal G}_I^\delta$ from below because for any $\gamma,p$ values, there are values of $\lambda $ such that ${\cal G}_I^{\delta}\geq {\cal H}_I^\delta$ holds. Since $q(T)\geq 0$ for ${}^\forall T\in {\rm supp}(q)$ and $Q(T)$ is a primitive function of $q(T)$, $Q(T)$ takes its minimum at $T=\tlb$, and thus, the first intersection of ${\cal G}_I^\delta$ and ${\cal H}_I^\delta$ is formed at $\lambda=\tlb$, and moves continuously with changes of the parameter values. Thus, the condition is sufficient. 

Especially, in the case of $\tlb=0$, the transition becomes continuous. By taking the contraposition of the argument,  the condition $\tlb>0$ is now shown as necessary.

\end{widetext}

\appendix
\section{Relaxing the assumption}\label{sec:relax_assumption}
In this section, we relax the assumption that the growth and death take place only in the single, growing state. We first specifically see what happens if the cells in the dormant state are killed by the antibiotics. Next, we study the model where the cells recover their growth and death rate gradually. 

\subsection{Non-zero killing rate at the full dormant state}
Here we introduce the non-zero killing rate also for the full dormant state. We assume that the killing rate at the full dormant state is smaller than that of the active state, and thus, we set it to $\alpha\gamma$ where $0<\alpha<1$. The rate equation is given as
\begin{eqnarray}
\dot{d}(t)&=&
\begin{cases}
-(\alpha\gamma+1/\lam)d(t) & (t<T)\\
-d(t)/\lam & (t>T)
\end{cases}
\\
\dot{g}(t)&=&
\begin{cases}
d(t)/\lam-\gamma g(t) & (t<T)\\
d(t)/\lam+g(t) & (t>T)\\
\end{cases}
\end{eqnarray}
The solutions of the equations and the fitness function are given by

\begin{figure}[htbp]
\begin{center}
\includegraphics[width = 85 mm, angle = 0]{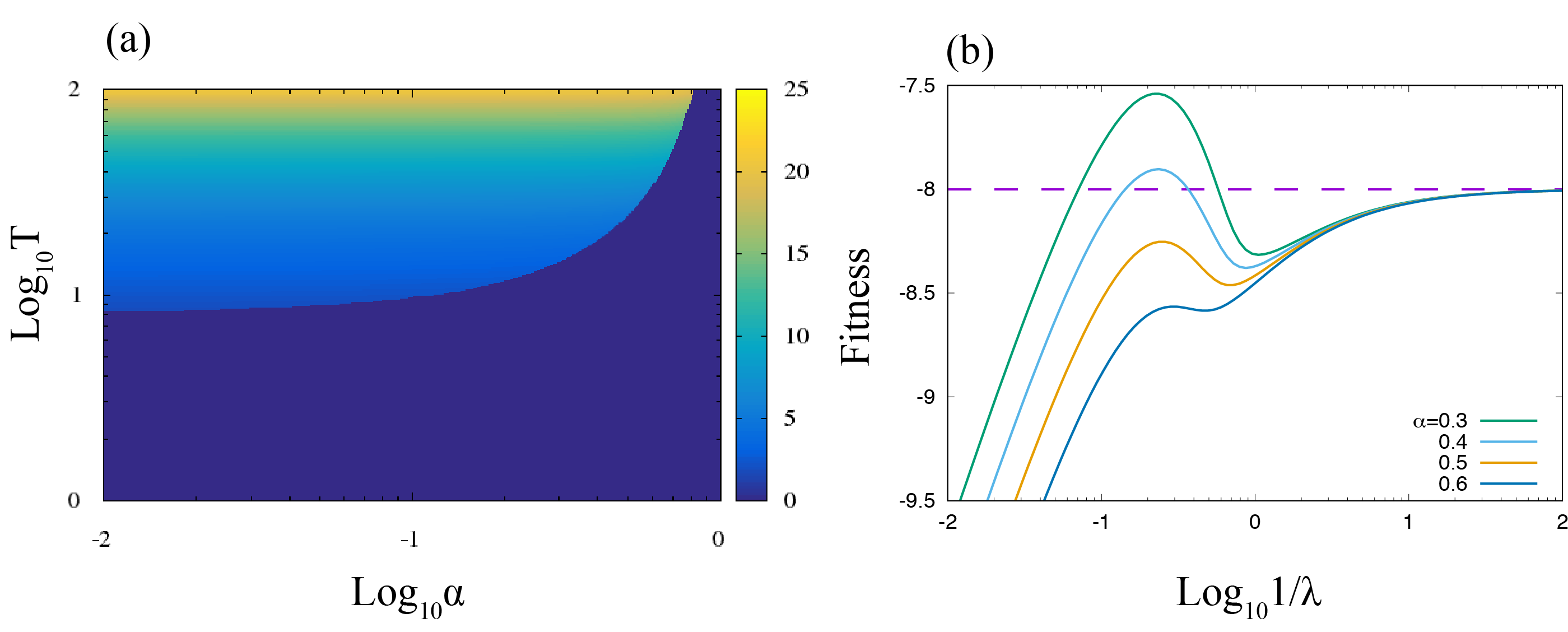}
\caption{(a). The optimal $\lam$ is plotted as a function of $\alpha$ and $T$. $\lam^*$ shows the discontinuous transition. (b). The fitness function for several choices of $\alpha$. $\lam^*$ transits in the same way as Fig1(b) in the main text. The dashed line represents $\tilde{F}_I(\lam=0,\alpha,\gamma,p,T)$. $p=0.2$, $\gamma=1.0$, and $T=20$ for (b).}
    \label{fig:alpha}
  \end{center}
\end{figure}

\begin{widetext}
\begin{eqnarray*}
d(t)&=&
\begin{cases}
e^{-(1/\lam+\alpha\gamma)t} &(t<T)\\
e^{-\alpha\gamma T}e^{-t/\lam} &(t>T)\\
\end{cases}\\
g(t)&=&
\begin{cases}
\frac{1}{1-(1-\alpha)\gamma\lam}\Bigl(e^{-\gamma t}-e^{-(1/\lam+\alpha\gamma)t}\Bigr) & (t<T)\\
\frac{1}{1-(1-\alpha)\gamma\lam}\Bigl(e^{-(1+\gamma )T}-e^{-(1+1/\lam+\alpha\gamma)T}\Bigr)e^t+\frac{1}{1+\lam}\Bigl(e^{-(1+1/\lam)T}e^t-e^{-t/\lam}\Bigr)e^{-\alpha\gamma T} &(t>T)
\end{cases}\\
\tilde{F}_I(\lam,\alpha,\gamma,p,T)&=&(1-p)\ln\Biggl[\frac{1}{1+\lam}\Biggr]+p\Biggl(-T+\ln\Biggl[\frac{1}{1-(1-\alpha)\gamma\lam}\Bigl(e^{-\gamma T}-e^{-(1/\lam+\alpha\gamma)T}\Bigr)+\frac{1}{1+\lam}e^{-(1/\lam+\alpha\gamma)T}\Biggl]\Biggr).
\end{eqnarray*}
\end{widetext}
Fig.\ref{fig:alpha}(a) shows the optimal lag time as the function of $\alpha$ and $T$. The optimal $\lam$ value still shows the transition by changing $T$. Since the nature of the transition is the same with $\alpha=0$ case, the transition is triggered by changing the severeness of the antibiotics application by changing either $p$, $\gamma$, and $T$. Also the model shows transition by changing $\alpha$ in the region $T\geqapprox 10$. The transition takes place discontinuously as shown in Fig.\ref{fig:alpha}(b). Note that the optimal lag time for $\alpha=1$ case is zero regardless of the other parameter values because there is no reason to stay at the dormant state. 

\subsection{A gradual growth resurrection model}
We relax the assumption that cells recover the ability for growth together with death abruptly when they reached to the single growing state. To making a gradual recovery of the growth and death rate possible, we use the model with multiple states. 

Here we consider the model has $M+1$ states ($M$ steps). Each state has the growth, death, and transition rate to next state, $\mu_i,\gamma_i,$ and $1/\lam_i$. Since the scope of this section is an impact of the gradual recovery of the growth and the death rate on the main result, here we assume that $\lam_i$'s are uniform among all the states while it is zero at the final, the $M$th state. Then, the temporal evolution of the population after an inoculation is ruled by 
\begin{eqnarray*}
\frac{d}{dt}N_i(t)&=&\bigl(\hat{\delta}_{i,0}N_{i-1}-\hat{\delta}_{i,M}N_i\bigr)M/\lambda+\alpha_i(t) N_i, \ (0\leq i\leq M)\\
\alpha_i(t)&=&
\begin{cases}
-\gamma_i\ & (t<T)\\
\mu_i\ &(t\geq T)
\end{cases}
\end{eqnarray*}
where $N_i$ represents the population of the cells at the $i$th state and  $\hat{\delta}_{i,j}$ is the complementary of the Kronecker's delta defined as $1-\delta_{i,j}$. We set $\mu_M$ as unity. The solution of the ordinary differential equation is given by

\begin{eqnarray}
N_i(t)&=&\sum_{j=0}^i\xi_{ij}c_je^{\beta_jtM/\lambda}\label{eq:Baranyi_Sol}
\end{eqnarray}where, parameters are given by 
\begin{eqnarray*}
\xi_{ij}&=&
\begin{cases}
\prod_{k=j+1}^i(\beta_j-\beta_k)^{-1},&(i>j),\\
1 &(i=j)\\
\end{cases}\\
\beta_j&=&\alpha_jM\lambda-\hat{\delta}_{j,M},
\end{eqnarray*}
and $\xi_{ij}'$s with $i<j$ do not appear in Eq.(\ref{eq:Baranyi_Sol}). We put a superscript $\pm$ to $N_i,\xi_{ij},\alpha_i,\beta_i$, and $c_i$ representing before($-$) and after ($+$) finishing the antibiotics application.  $c_i^\pm{}'$s are determined to satisfy the initial condition  $N_0^-(0)=1,N_1^-(0)=\cdots =N_{M}^-(0)=0$ and continuity of the solution at the end of the antibiotics application, $N_i^-(T)=N_i^+(T)$. 

\begin{figure}[htbp]
\begin{center}
\includegraphics[width = 85 mm, angle = 0]{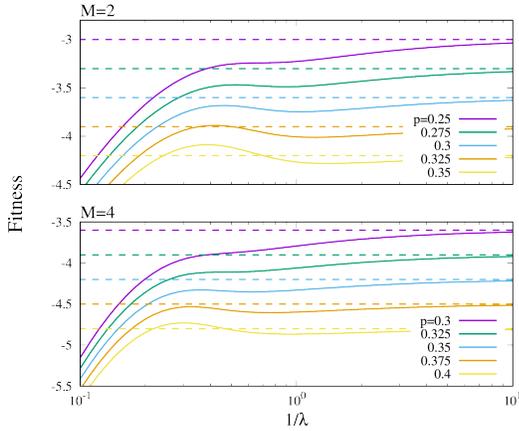}
\caption{The fitness values of the $M$ steps, gradual resurrection model are plotted against $\lambda$ for several values of $p$ where $M=2$ (top) and $4$ (bottom). The dashed lines represents the fitness value at $\lambda=0$ of the corresponding color. As $p$ value increases, the fitness function forms a peak and the value at the peak exceeds the value at $\lambda=0$. Here $\mu_i$ and $\gamma_i$ are given as $\mu_i=\gamma_i=i/M,\ (0\leq n\leq M)$, and $T=6.0$.}
    \label{fig:gradual}
  \end{center}
\end{figure}

The antibiotics-free solution is obtained by setting $T=0$. Since the definition of the fitness of a single round is the logarithmic growth of the population under a large $t$ limit, relative to the zero- antibiotics application time and lag time, it is given as
$$f(T)=\lim_{t\to \infty}\frac{\sum_{i=0}^{M}\sum_{j=0}^i\xi_{ij}^+c_j^+e^{\beta_j^+ tM/\lambda}}{e^t}=c_{M}^+,$$
and accordingly, the fitness function is calculated. The fitness function is plotted against $\lambda$ for several choices of $p$ values and $M=2$ and $4$ in Fig.\ref{fig:gradual}. The fitness function has its maximum at $\lambda=0$ when $p$ is small, while it forms local maximal which exceeds the fitness value at the origin as $p$ increases. The result suggests the robustness of the discontinuous transition described in the main text against the model extension that the cells recover their growth rate and antibiotic susceptibility.

\section{The threshold of the extinction}\label{sec:extinction}
Here we study the slightly modified model in which the population cannot be lower than a threshold value. Note that the populations ($d(t)$ and $g(t)$ in Eq.(\ref{eq:dynamics})) decrease only in the duration of the antibiotics application, and the shorter the lag time $\lambda$ is, the more prominent the killing effect gets. If $\lambda$ is sufficiently larger than $T$, the total population $d(t)+g(t)$ never reaches to the threshold. Therefore, the introduction of the threshold sets the lower limit of the lag time $\lambda$ so that all the cells are not killed by the antibiotics. 

To guarantee the population is greater than a threshold denoted by $\delta_{\rm ext}$, the inequality 
\begin{equation}
d(T)+g(T)=\frac{\lambda\gamma}{\lambda\gamma-1}e^{-T/\lambda}-\frac{1}{\lambda\gamma-1}e^{-\gamma T}>\delta_{\rm ext} \label{eq:thresh}
\end{equation}
needs to be satisfied. We set $\delta_{\rm ext}=10^{-6}$ because the initial value of $(d
+g)$ is normalized to unity in the main text and typical numbers of the bacteria in a test tube for a few milliliters range from $10^6$ to $10^9$. By substituting $\lambda=0$ into Eq.(\ref{eq:thresh}), we get the condition for allowing zero lag time, given as  $e^{-\gamma T}>\delta_{\rm ext}$. If this condition is not satisfied, the minimum lag time $\lam_{\rm min}$ is set by Eq.(\ref{eq:thresh}) with the replacement of inequality with equality. 

\begin{figure}[htbp]
\begin{center}
\includegraphics[width = 85 mm, angle = 0]{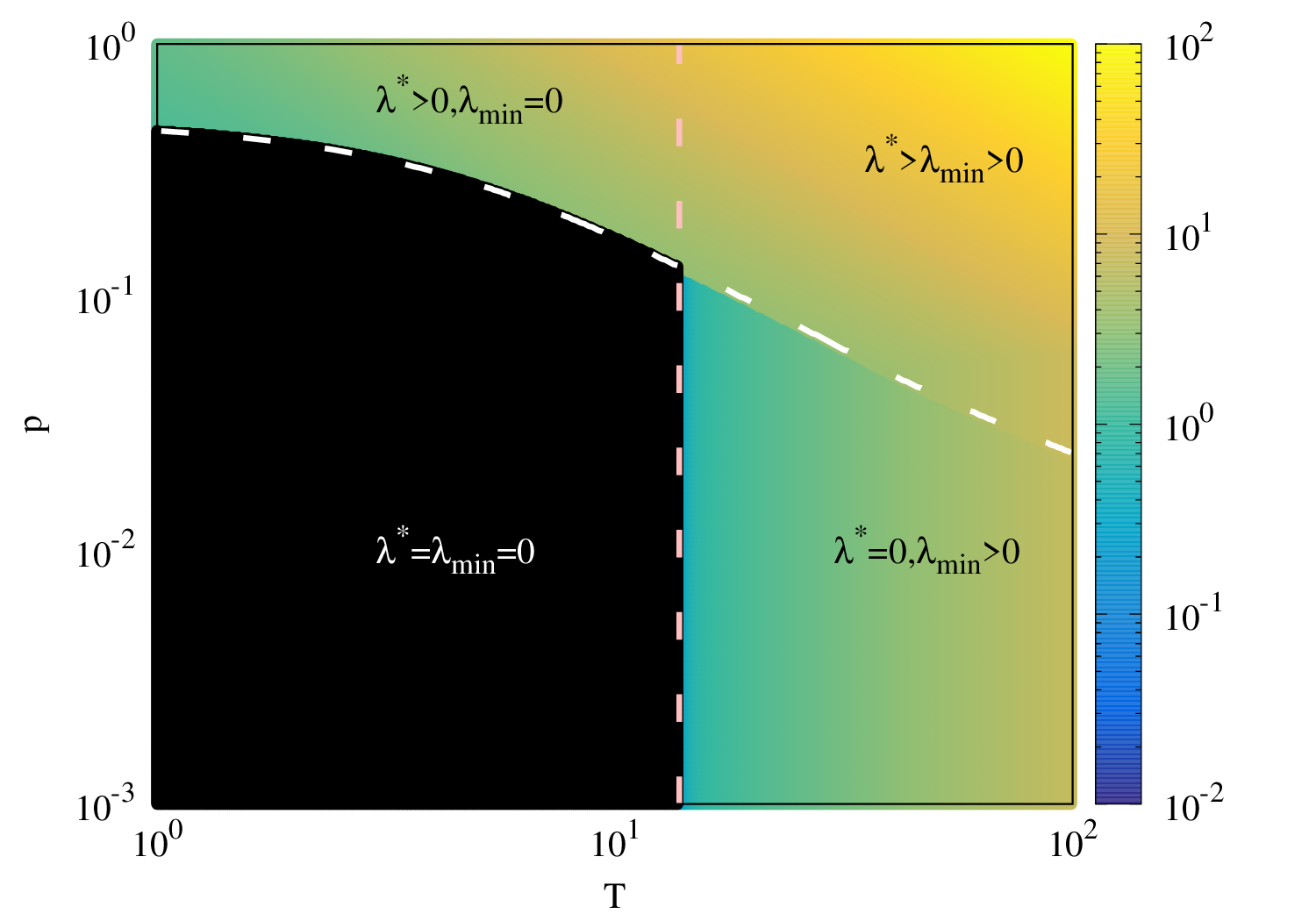}
\caption{The heatmap of the optimal lag time. The white and pink dashed lines divide the space into four regions. Color shows ${\rm max}\{\lam^*,\lam_{\rm min}\}$.  $\gamma$ is set to unity.}
    \label{fig:lam_min}
  \end{center}
\end{figure}

Fig.\ref{fig:lam_min} shows ${\rm max}\{\lam^*,\lam_{\rm min}\}$ in $(T,p)$ plane where $\lambda^*$ denotes the optimal average lag time without the restriction (Eq.(\ref{eq:thresh})). The plane is divided into four regions: (i). the zero lag time is optimal and feasible, (ii). the zero lag time is optimal but $\lam_{\rm min}$ is non-zero, (iii).  $\lam_{\rm min}$ is non-zero but the optimal lag time is greater than that, and (iv). the optimal lag time is non-zero and $\lam_{\rm min}=0$. The white line gives the boundary between $\lam^*=0$ and $\lam^*>0$, and the pink line being given as $-\ln(\delta_{\rm ext})/\gamma$ divides the region into $\lam_{\rm min}=0$ and $\lam_{\rm min}>0$ part.

The figure tells that, by changing $p$ as the parameter, one can observe the discontinuous transition for any reasonable $T$ value while the jump size of the optimal lag time gets less as $T$ becomes larger. On the other hand, when $T$ is chosen as the parameter, one can observe only the continuous transition in a reasonable timescale with a rare antibiotics application ($p$ as less than a few percent).

\section{The equation used for the evolution simulation of the lag time}\label{sec:evolution}
The system of equations used for the evolution simulation (in the section III-C) is following
\begin{widetext}
\begin{eqnarray*}
\frac{d}{dt}d^{(i)}(t)&=&-d^{(i)}(t)/\lambda_i\ (0\leq i<N)\\
\frac{d}{dt}g^{(0)}(t)&=&d^{(0)}(t)/\lambda_0+(1-\epsilon)\alpha g^{(0)}(t)+\epsilon \alpha g^{(1)}(t)\\
\frac{d}{dt}g^{(i)}(t)&=&d^{(i)}(t)/\lambda_i+(1-2\epsilon)\alpha g^{(i)}(t)+\epsilon \alpha (g^{(i-1)}(t)+g^{(i+1)}(t)),\ (1\leq i\leq N-2)\\
\frac{d}{dt}g^{(N-1)}(t)&=&d^{(N-1)}(t)/\lambda_{N-1}+(1-\epsilon) \alpha g^{(N-1)}(t)+\epsilon \alpha g^{(N-2)}(t),
\end{eqnarray*}
\end{widetext}
where $d^{(i)}$ and $g^{(i)}$ represents the population of the $i$th-type cells in the dormant, and the growing state, respectively. $(\alpha,\epsilon)$ is $(-\gamma,0)$ if the antibiotics is applied and $t<T$, otherwise, it is $(1.0,10^{-3})$.

\section{A multi-step model}\label{sec:multistep}
As an extension of the model (Eq.(\ref{eq:dynamics})-(\ref{eq:dynamics2})), we consider the model with multiple dormant states. All the cells start to wake up from the $0$th dormant state after the re-inoculation, and transit one by one through the $M$ dormant states to regain the growth rate at the growing state. Here, we assume that the growth rate and the death rate are non-zero only in the growing state, and the transition rates from one state to the next state are uniform (this is a special case of the gradual resurrection model). Then, the temporal evolution of the population after an inoculation is ruled by 
\begin{eqnarray}
\frac{d}{dt}d_0(t)&=&-d_0(t)M/\lambda,\\ 
\frac{d}{dt}d_i(t)&=&(d_{i-1}-d_i(t))M/\lambda, \ (1\leq i \leq M-1)\\ 
\frac{d}{dt}g(t)&=&
\begin{cases}
d_{M-1}(t)M/\lambda-\gamma g(t) & (t<T)\\
d_{M-1}(t)M/\lambda+ g(t) & (t>T),\\
\end{cases}
\end{eqnarray}
where $M/\lam$ is the rate of the transition from the $i$th to the $i+1$th state. 

This dynamics leads to Erlang distribution 
$P(l)=l^M e^{-l/ \lambda_M}/(M!\lambda_M^{M+1})$ as the lag time distribution where $\lambda_M=\lambda/M$. The average lag time is $\lambda$. The single-round fitness $f_M(T)$ is now given by 
\begin{widetext}
\begin{eqnarray}
f_M(T)&=&e^{-(1+\gamma)T}(1-\gamma\lambda_M)^{-(1+M)}+e^{-(1+1/\lambda_M)T}\sum_{n=1}^{M+1}\frac{(T/\lambda_M)^{M+1-n}}{(M+1-n)!}\Bigl((1+\lambda_M)^{-n}-(1-\gamma\lambda_M)^{-n}\Bigr).\label{eq:Erlang}
\end{eqnarray}
\end{widetext}

As described in {\it Supplementary Information} section II, $M$-step models, including the main model ($M=1$ case) in the manuscript, generally exhibit the discontinuous transition of $\lambda$ as long as the lower bound of the support of $q(T)$, a distribution of the antibiotics application time, is non-zero. 
It is worth mentioning that as $M$ increases, the optimal fitness value becomes greater if $T$ is fixed because the Erlang distribution with a large $M$ is peakier. However, if $T$ is distributed, a large $M$ is not always beneficial, and there typically optimal $M$ exists.

\section{The computational procedure for obtaining the optimal combined-Erlang distributions}
\label{sec:opt_Erlang}
For obtaining the optimal lag time distribution of the sequential model with $N_p$ phenotypes and $M$ steps (Fig.\ref{fig:opt_erlang}), we carried out following computations. 

For given $N_p$, $M$, and other parameters ($p, \gamma$, and $q(T)$), the fitness value 
$$F=\int_0^{\rm T{max}}\hat{q}(T)\ln\Bigl[\sum_{i=0}^{N_p-1}x_if_M(T;\lambda_i)\Bigr]dT,$$
is maximized where $\hat{q}(T)=(1-p)\delta(T)+pq(T)$. $f_M(T;\lambda_i)$ is given in Eq.(\ref{eq:Erlang}) while $\lambda$ is replaced  by $\lambda_i$, and $x_i$ represents the fraction of the $i$th phenotype. Here, the variables for the optimization are $x_i'$s and $\lambda_i'$s and the restriction $\sum_{i=0}^{N_p-1}x_i=1$ is introduced.

\section{Mapping to the Kussell-Leibler model}\label{sec:KL}
The focus of the paper written by Kussell and Leibler \cite{kussell2005phenotypic} is to ask the optimal strategy of the bacterial population which has multiple types (either genotypic or phenotypic) to grow most efficiently under the fluctuating environment. The interactions among cells are not considered in the model, and thus, the model is given as the linear ordinary differential equations, $\dot{\bf{x}}(t)=A_k{\bf x}(t)$ where ${\bf x}$ represents the population vector at time $t$. The matrix $A_k$ consists of the growth term, and in addition, the transition term among the types representing either the genetic mutation or phenotypic switch in the environment $k$. The matrix changes from $k$ to $k'$ $(k\neq k')$ after a certain period of time has passed to emulate the fluctuations of the environment. The growth rate of the $i$th species (whose population is represented by the $i$th column of ${\bf x}$) has a different value depending on the environment index $k$. It can be either positive or negative. The negative growth rate then indicates that the $i$th type cannot even survive in the environment $k$ and dies over time. Because of the environmental dependency of the growth rate, the population needs to tune the transition rate among the types so that the total number of the cells increases well.

The authors studied two scenarios, namely the responsible adaptation and the stochastic adaptation. In the responsible adaptation, the cells can "sense" the environment and use the environment-specific transition rates. In contrast, in the stochastic adaptation, the cells are allowed to use only a set of transition rates independently of the environment. One of the main results of the paper is that the optimal strategy for stochastic adaptation is mimicking the fluctuation of the environment. For instance, if the number of the types and the number of environments are the same and the $i$th type has the highest growth rate in the $i$th environment, the optimal transition rate of the $i$th to the $j$th species, $H(i\to j)$ is given as $b(i\to j)/\tau_i$ where $b(i\to j)$ is the transition rate from the $i$th to $j$th environment, and $\tau_i$ represents the latency time of the $i$th environment. The optimal lag time distribution is not the complete copy of the distribution of the antibiotics application time, and thus, there should be mathematical differences between the KL model and our model.

To show how to map our model to this framework and where the different consequence comes from, we briefly explain the analysis done by the authors. Since the solution of the ordinary differential equation is given as 
${\bf x}(t)=\exp[A_k t]{\bf x}(0)$, we obtain a sequence 
\begin{eqnarray*}
{\bf x}_{\rm ini}&\to& e^{A_{\mathcal{E}_0}\tau}{\bf x}_{\rm ini}\to e^{A_{\mathcal{E}_1}\tau} e^{A_{\mathcal{E}_0}\tau}{\bf x}_{\rm ini}\to\cdots\\
&\to&\prod_{i=0}^{N} e^{A_{\mathcal{E}_i}\tau}{\bf x}_{\rm ini}\to\cdots
\end{eqnarray*}
representing the total population just before the $i$th change of the environmental condition, where ${\mathcal E}_i$ represents the environment at the $i$th round. Here we eliminated the environment dependency of the latency time $\tau$ because the dependency is unnecessary for the later arguments as long as $\tau$ is sufficiently large.

The analysis in the paper is calculating the average Lyapunov exponent over the sequential environmental changes and maximizing it by modulating the transition rate among the types. Thus, by denoting the population vector after the $n$th round as ${\bf x}_n$, the question is formulated as the optimization problem of 
\begin{equation}
    F_{KL}=\Biggl\langle \frac{{\bf 1}\cdot e^{A_{\mathcal{E}_{n+1}}\tau}{\bf x}_n}{{\bf 1}\cdot {\bf x}_n}\Biggr\rangle \label{eq:KL_fitness}
    \end{equation}
where the average $\langle\cdot\rangle$ is taken over $n$'s and ${\bf 1}$ is the vector having the same dimension with ${\bf x}_n$ given as ${\bf 1}={}^t(1,1,\cdots,1)$. The optimal transition rates among the types are calculated by using the perturbation method. The population growth in a single round is determined only from the growth rates, while the transition rates play a role in modulating the population distribution among the types. The central assumption is that there is a separation among the eigenvalues of $A_k$ to simplify the evaluation of the fitness in a single round. If the assumption holds, the fitness is well-approximated by the highest growth rate in the environment $k$ and the population of the fittest type under the large $\tau$ limit. When the assumption is violated, the contribution of all the other types are unignorable, and thus, the simple argument is no longer valid.

Now, we describe how to map our model to this framework. We consider the generic case that the population of the cells has a lag time distribution $r(l)$. We set the upper limit of the distribution $T_{\rm max}$ and discretize the distribution by a certain size of the bin to make a finite number of subgroups having the lag time $l_i$. Now, the different "types" in the KL framework corresponds to the groups of the cells with different lag time and the different environment means the different antibiotics application time $T_k$. We suppose that the number of types and the number of the environment are equal for simplicity. 

Our population dynamics is non-autonomous because the cells of the $i$th type are in quiescent until $t=l_i$. However, as long as $T_{\rm max}<\tau$ holds, we can write down the autonomous, effective population dynamics given as 
\begin{equation}
\dot{{\bf y}}(t)=B_k{\bf y}(t)\label{eq:KLmap}
\end{equation}
with
\begin{eqnarray*}
B_k=E-{\rm diag}\Bigl(\ln[ f(l_0,T_k)]/\tau,\cdots,\ln[f(l_{N-1},T_k)]/\tau\Bigr)
\end{eqnarray*}
where $E$ is the unit matrix and ${\bf y}(t)$ is the population vector at time $t$ and $\ln f(l_i,T_k)$ represents the population loss by having the lag time $l_i$ and exposed to the antibiotics for $T_k$. Note that the number of the $i$th type cells at time $\tau$ is given as $f(l_i,T_k)\exp[\tau]y_i(0)$ being consistent with the original population dynamics in the main text. Now, we can emulate our population dynamics in the KL framework by choosing the discretized lag time distribution ${\bf r}$ as the initial vector ${\bf y}$(0). Note that Eq.(\ref{eq:KL_fitness}) is defined for an arbitrary population vector ${\bf x}_n$, and thus, starting from the same vector ${\bf r}$ every time is a special case of Eq.(\ref{eq:KL_fitness}). Also, Eq.(\ref{eq:KL_fitness}) is invariant for the normalization of the total population at every round. 

Now, we mapped the present model to the KL framework, and thus, are able to see where the difference between the prediction in ref.~\cite{kussell2005phenotypic} and the present result comes from. The critical difference is that the separation of the eigenvalue never happens under $\tau\to\infty$ limit in the present model. The growth rate correspondence in the present model, $1-f(l_i,T_k)/\tau$ lose the $i$-dependence under this limit. Thus, there is no way to evaluate the population growth by only a single eigenvector, and it leads to a different consequence.

\bibliography{ref.bib}

\end{document}